\documentclass{article}

\usepackage{arxiv}

\usepackage[utf8]{inputenc} 
\usepackage[T1]{fontenc}    
\usepackage{hyperref}       
\usepackage{url}            
\usepackage{booktabs}       
\usepackage{amsfonts}       
\usepackage{nicefrac}       
\usepackage{microtype}      
\usepackage{lipsum}		
\usepackage{graphicx}
\usepackage{natbib}
\usepackage{doi}
\usepackage{newtxtext,newtxmath}
\usepackage{aas_macros}
\usepackage{amsmath}	
\usepackage[utf8]{inputenc}

\newcommand{\dif}{\mbox{d}}

\title{Collisional properties of cm-sized high-porosity ice and dust aggregates and their applications to early planet formation}

\author{ \href{https://orcid.org/0000-0001-9467-7685}{\includegraphics[scale=0.06]{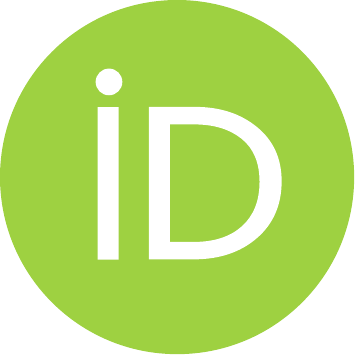}\hspace{1mm}Rainer R.~Schr\"apler} \\
	Institut f\"ur Geophysik und extraterrestrische Physik\\
	Technische Universit\"at Braunschweig\\
	Mendelssohnstr. 3,D-38100 Braunschweig, Germany \\
	\texttt{r.schraepler@tu-braunschweig.de} \\
	\And
	
	\hspace{1mm}Wolf A.~Landeck \\
	Institut f\"ur Geophysik und extraterrestrische Physik\\
	Technische Universit\"at Braunschweig\\
	Mendelssohnstr. 3,D-38100 Braunschweig, Germany \\
	\texttt{a.landeck@tu-braunschweig.de}\\
	
	\And
	\href{https://orcid.org/0000-0003-1531-737X}{\includegraphics[scale=0.06]{orcid.pdf}\hspace{1mm}J\"urgen~Blum} \\
	Institut f\"ur Geophysik und extraterrestrische Physik\\
	Technische Universit\"at Braunschweig\\
	Mendelssohnstr. 3,D-38100 Braunschweig, Germany \\
	\texttt{j.blum@tu-braunschweig.de}\\
	
}

\renewcommand{\headeright}{A preprint (accepted by MNRAS)}
\renewcommand{\undertitle}{A preprint (accepted by MNRAS)}
\renewcommand{\shorttitle}{\textit{arXiv} Template}

\hypersetup{
pdftitle={A template for the arxiv style},
pdfsubject={q-bio.NC, q-bio.QM},
pdfauthor={David S.~Hippocampus, Elias D.~Striatum},
pdfkeywords={First keyword, Second keyword, More},
}

\begin{document}
\maketitle

\begin{abstract}
In dead zones of protoplanetary discs, it is assumed that micrometre-sized particles grow Brownian, sediment to the midplane and drift radially inward. When collisional compaction sets in, the growing aggregates collect slower and therefore dynamically smaller particles. This sedimentation and growth phase of highly porous ice and dust aggregates is simulated with laboratory experiments in which we obtained mm- to cm-sized ice aggregates with a porosity of 90\% as well as cm-sized dust agglomerates with a porosity of 85\%. We modelled the growth process during sedimentation in an analytical calculation to compute the agglomerate sizes when they reach the midplane of the protoplanetary disc. In the midplane, the dust particles form a thin dense layer and gain relative velocities by, e.g., the streaming instability or the onset of shear turbulence. To investigate also these collisions, we performed additional laboratory drop tower experiments with the high-porosity aggregates formed in the sedimentary-growth experiments and determined their mechanical parameters, including their sticking threshold velocity, which is important for their further collisional evolution on their way to form planetesimals. Finally, we developed a method to calculate the packing-density-dependent fundamental properties of our dust and ice agglomerates, the Young's modulus, the Poisson ratio, the shear viscosity and the bulk viscosity from compression measurements. With these parameters, it was possible to derive the coefficient of restitution which fits our measurements. In order to physically describe these outcomes, we applied a collision model. With this model, predictions about general dust-aggregate collisions are possible.
\end{abstract}


\section{Introduction}\label{cec:INTRO}
In the accretion phase of a protoplanetary disc (PPD), most of the solid material is evaporated when passing though the accretion shock \citep{DraineSalpeter}. Following this phase, it is widely assumed that dusty material condenses to sub-micrometre- to micrometre-sized dust particles in the inner part of the disc. For example \citet{Kimura2009} gained in his dust-condensation experiments particle sizes of 500 nm. Beyond the snow line, in the outer parts of the disc, ice particles condense to similar sizes. In the early phase of the subsequent coagulation phase, these particles grow by  Brownian-motion-driven hit-and-stick collisions to fractal agglomerates \citep{Blumetal2000,Krause2004,Blum2006}. From a given size on, the particles get collisionally compacted \citep{DominikTielens1997,BlumWurm2000}. In a flared disc on the surface of the dead zone, particles are compacted by intruding turbulent motions \citep{Ruff1987}, photophoresis \citep{Matthews2016} and photoevaporation, the latter only for ice particles \citep{Bodenstein2018}. 

Compacted particles sediment faster than fluffy ones and collect less dense particles on their way to the midplane. We simulate this sedimentary growth process in laboratory experiments by coupling micrometre-sized monomer particles to a gas flow, which transports them to a target (see Sects. \ref{sect:cuboids} and \ref{sect:RBDice}), as well as by an analytical model (see Section \ref{sect:SEDI}). Due to the sedimentation process, the dust particle concentration in the midplane increases until a dust-dominated sub-disc forms. This sub-disc causes shear turbulence  \citep{Dubrulle,Cuzzi1993}, which in turn leads to increased relative velocities between the dust particles \citep{SchrHen04} and Section \ref{sect:CSD}. To understand the collisional evolution in this growth phase, we  performed collision experiments with our high-porosity dust and ice agglomerates using laboratory drop towers, as described in Section \ref{sect:EXP}. The results of these experiments provide the sticking thresholds and the coefficients of restitution of the fluffy agglomerates.

Finally, we also can answer the fundamental question whether bouncing occurs also for high-porosity dust or ice aggregates. Based on numerical simulations of aggregate collisions, \citet{wadaetal2011} had suggested that bouncing as an intermediate step between sticking (at low velocities) and fragmentation (at high velocities) does not occur if the mean coordination number, i.e. the number of contacting neighbours of a dust monomer inside an aggregate, is smaller than 6. 

\section{Growth of agglomerates during the laminar sedimentation towards the midplane }\label{sect:SEDI}
Before we present our experimental results in Sect. \ref{sect:RBD}, we will first consider the aggregate growth due to and during the sedimentation motion in a PPD. 

\subsection{How large can dust aggregates grow by collisional sticking in the sedimentation phase?}\label{sect:sizeh}
To estimate the agglomerate size in the midplane of a protoplanetary disc, we compute an analytical sedimentation model. We start with a particle infinitely high ($h=\infty$) above the disc's midplane ($h=0$) that has a bulk density of $\rho_\mathrm{B}$. On its way down towards the midplane through a disc with a local spatial dust density of $\rho(h)$, the particle collects all grains (which we assume to be stationary) it geometrically encounters, and grows by the ballistic particle-cluster-aggregation (BPCA) process with a sticking probability of unity to a radius $r$. The BPCA process causes a fractal dimension of the growing aggregates of 3 \citep{mukaietal1992}, i.e. the bulk density of the aggregates and their internal packing fraction (or volume filling factor) $\Phi$ are constant. Thus, the growing aggregate will gradually decouple from its surrounding gas while it continues to sediment. 

\begin{figure}
    \center
    \includegraphics[width=0.7\textwidth]{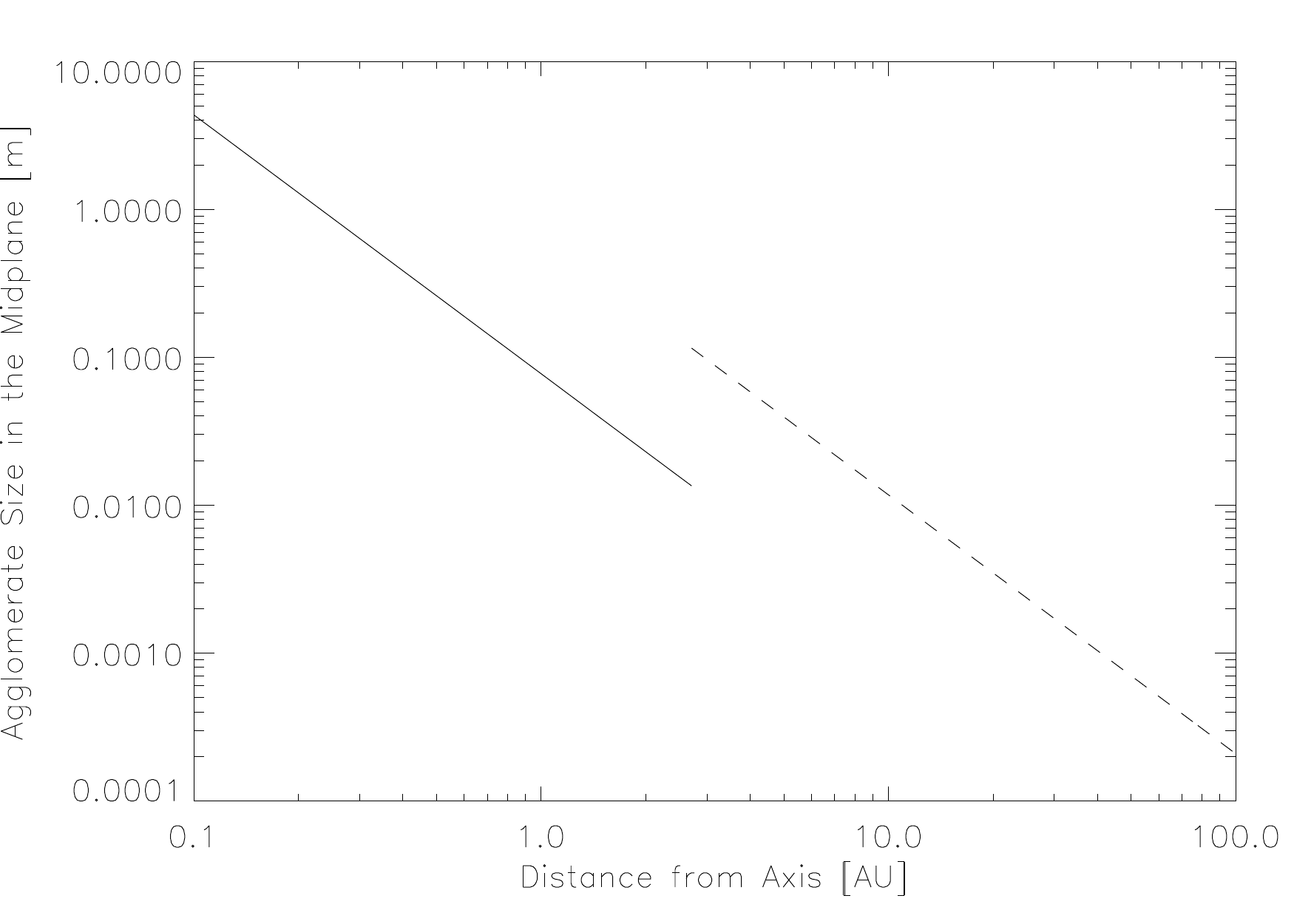}
    \caption{\label{fig:PebSiz} The agglomerate radius of the aggregates when they reach the midplane of the PPD as a function of the distance from the disc centre. Inside the snow line at 2.7 au, it is assumed that the disc contains only dust and gas with a mass ratio of 1:100 (solid line). Outside the snow line, it is assumed that the disc contains dust and ice (dashed line) with a solid-to-gas mass ratio of 1:33.}
 \end{figure}

On its way down by a path length of $\dif h$, the aggregate will collect a dust mass of 
\begin{equation}
\dif m = -r^2 \pi \rho(h) \dif h,
\end{equation}
with
\begin{equation}
m=\frac43 \pi  r^3 \Phi \rho_\mathrm{S} ,
\end{equation}
with the mass density of the solid grains $\rho_\mathrm{S}$ and $\rho_\mathrm{B}=\Phi \rho_\mathrm{S}$, which implies
\begin{equation}
\dif m=4 \pi r^2 \Phi \rho_\mathrm{S} \dif r
\end{equation}
and 
\begin{equation}
\dif r=-\frac{1}{4}\frac{\rho(h)}{\Phi \rho_\mathrm{S}} \dif h.
\end{equation}
Using the vertical dust density distribution \citep[e.g.][]{Dubrulle}
\begin{equation}
\rho(h)=\rho_0 \mbox{Exp}\left(-\frac{h^2}{H^2}\right),
\end{equation}
where $H$ and $\rho_0$ are the scale height of the gas and the dust density at the midplane at a given distance $R$ from the disc centre (see Table \ref{table:Para}), respectively. The agglomerate radius at the midplane is then
\begin{equation}
    r=-\frac{\rho_0}{4 \Phi \rho_\mathrm{S}}\int_\infty^0 \mbox{Exp}\left(-\frac{h^2}{H^2}\right) \dif h,
\end{equation}
which yields

\begin{equation}
   r=\frac{\sqrt{\pi}}{2}\frac{\rho_0}{4 \Phi \rho_\mathrm{S}} H. 
\end{equation}
An overall result of this calculation is that the achieved agglomerate size is independent of the size of the monomer particles they consists of. We applied this finding to a minimum mass disc \citep[e.g.][]{Cuzzi1993} and used dust particles inside and a 3:1 ice-to-dust mixture outside the snow line located at 2.7 au. The used parameters are shown in Table \ref{table:Para}. We plotted the aggregate radius at the midplane as a function of the distance from the disc centre in Figure \ref{fig:PebSiz}. At 1 au, the final aggregate size is several cm; beyond 10 au, the aggregate radii do not exceed a few mm. 

\begin{table}
        \centering
        \caption{\label{table:Para} Assumed properties of the PPD as well as of the dust and ice particles. References: 1: this work; 2: \citet{BS2004}; 3: deduced from \citet{Cuzzi1993} with $R_\mathrm{D}$ and $\rho_{g,0}=1.4\times 10^{-6}~ \mathrm{kg ~ m^{-3}}$ being the distance from the disc centre and the mass density of the PPD gas in the midplane at 1 au, respectively.}
\begin{tabular}{l c c c c c c}
\hline
\hline
Parameter & Property & Reference \\
\hline
Material density $\rho_\mathrm{S}$ dust & 2.000 kg m$^{-3}$\\
Material density $\rho_\mathrm{S}$ ice & 930 kg m$^{-3}$\\
Material density $\rho_\mathrm{S}$ 3:1 ice-dust mixture & 1.197 kg m$^{-3}$\\
Packing density dust agglomerates $\Phi$ & 0.1 & 1\\
Packing density ice agglomerates $\Phi$ & 0.15 & 2\\
Packing density 3:1 dust-ice mixture $\Phi$ & 0.1125\\
Gas scale height & $H \approx 0.05 R$ & 3\\
Midplane gas density $\rho_g$ & $\rho_{g,0}\left(\frac{R_\mathrm{D}}{1 au}\right)^{-2.75}$ & 3\\
Dust-to-gas mass ratio & 0.01 & 3    \\
Ice-to-gas mass ratio & 0.03 &    \\
Snow line distance from disc center  & 2.7 au & \\
\hline
\end{tabular}
\end{table}

\subsection{Estimation of the particle sedimentation timescale} \label{sec:SedTime}

\begin{figure}
  \center
      \includegraphics[width=0.7\textwidth]{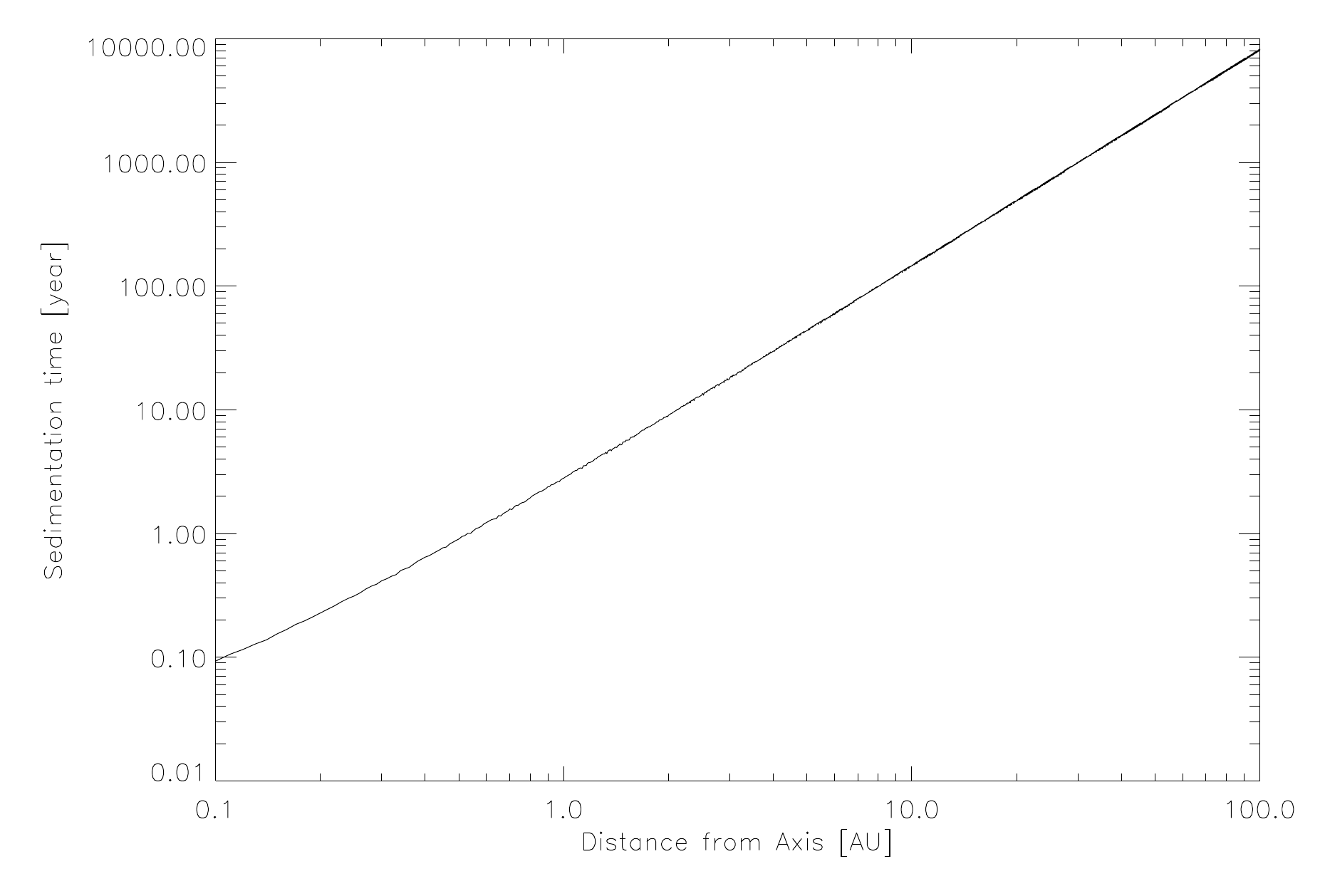}
  \caption{\label{fig:SediTime}  The timescale of the sedimentary-growth process as a function of the Kepler radius in a minimum mass PPD.}
 \end{figure}

To estimate the duration of the sedimentary-growth process described in the previous section, we apply the following approximation
\begin{equation}
    t=\int_{h_\mathrm{t}}^{h_\mathrm{b}}\frac{\dif h}{v(h)}\label{eq_t},
\end{equation}
where $v(h)$ is the sedimentation velocity at a height $h$ above the PPD midplane. As an integration from an infinite initial height to the midplane would lead to an infinite sedimentation time, we have to limit the origin of the path to a finite height. We chose the surface of the dead zone at $h_\mathrm{t}=1.5 H$ \citep[e.g.][]{Gole2016}, where $H$ denotes the gas scale height of the disc (see Table \ref{table:Para}). Because the sedimentation velocity at $h_\mathrm{b}=0$ is zero, we also have to determine the termination of the path. In case all particles would sediment to the midplane, the disc gets dust-dominated, which causes shear turbulence. This turbulence limits the final dust sub-disc to a height of $10^{-4} H$ \citep[e.g.][]{SchrHen04}. Therefore, we set the final height of the sedimentation process to $h_\mathrm{b}=10^{-4} H$. The sedimentation acceleration driven by the solar gravitation component in the vertical direction is
\begin{equation}
    g_h=\frac{g~h}{R},
\end{equation}
where $g$ is the gravitational acceleration of the central star at an orbital distance of $R$. The gas-grain friction time in the Epstein regime is
\begin{equation}
    \tau_f=\frac{\Phi \rho_\mathrm{S} r}{\rho_\mathrm{G} c},
\end{equation}
where $\rho_\mathrm{G}$ and $c$ are the mass density and the sound velocity of the gas. Hence, the sedimentation velocity is 
\begin{equation}
    v(h)=\tau_f g_h.
\end{equation}
The integration of Eq. \ref{eq_t} was performed numerically and its result as a function of distance to the centre of the disc is shown in Figure \ref{fig:SediTime}. The difference in sedimentation time for a dust-ice mixture (outside the snow line) and pure dust (inside the snow line) is very small and not visible in Figure \ref{fig:SediTime}. We found that the duration of the sedimentary-growth process is between 0.2 yrs at 0.1 au and 18,000 yrs at 100 au.

\subsection{The effect of grain charging on the growth scenario}

As shown by  \citet{Okuzumi2009}, grain charging can dramatically affect the growth of dust agglomerates. Moreover, \citet{Akimkin2020} performed numerical simulations on the electrostatic repulsion of grains in PPDs. Both works find a strong positive charging of dust grains from photoevaporation above the dead zone of the disc. Inside the dead zone, the grains obtain their charge from the weakly charged gas in which electrons are thermalised. This means that electrons are swept up by dust particles  until the electric energy in between the now charged dust particles and the free floating electrons equals the thermal energy. 
Because electrons as well as dust particles have the same thermal energy inside the dead zone, the dust particles cannot collide with their thermal velocity as well. Sedimenting particles, however, move to the midplane at much larger speed than that of thermal kinetic energies so that the sedimentary-growth process sketched above is not affected by grain charging. In the models of \citet{Akimkin2020}, strong grain growth appears in the transition region from the dead zone to the active zone. This is caused by turbulent mixing of the positively charged particles from the active zone with the negatively charged ones from the dead zone, which supports our proposed growth scenario, because the seeds that sediment can be produced there.  

\section{Experimental realisation of the sedimentary growth process}
\label{sect:RBD}
The sedimentary-growth scenario described in the previous section can relatively easily be approximated in laboratory experiments by letting individual monomer grains gently sediment in a rarefied gas atmosphere onto a growing target agglomerate \citep[see][]{BlumSchr2004,BlumSchr2006,Gundlachetal2011}. The corresponding laboratory methods and the properties of the forming aggregates will be described in the following section.

\subsection{Experimental technique}

\subsubsection{Silica aggregates}\label{sect:cuboids}
\begin{figure}
\center
      \includegraphics[width=0.5\textwidth]{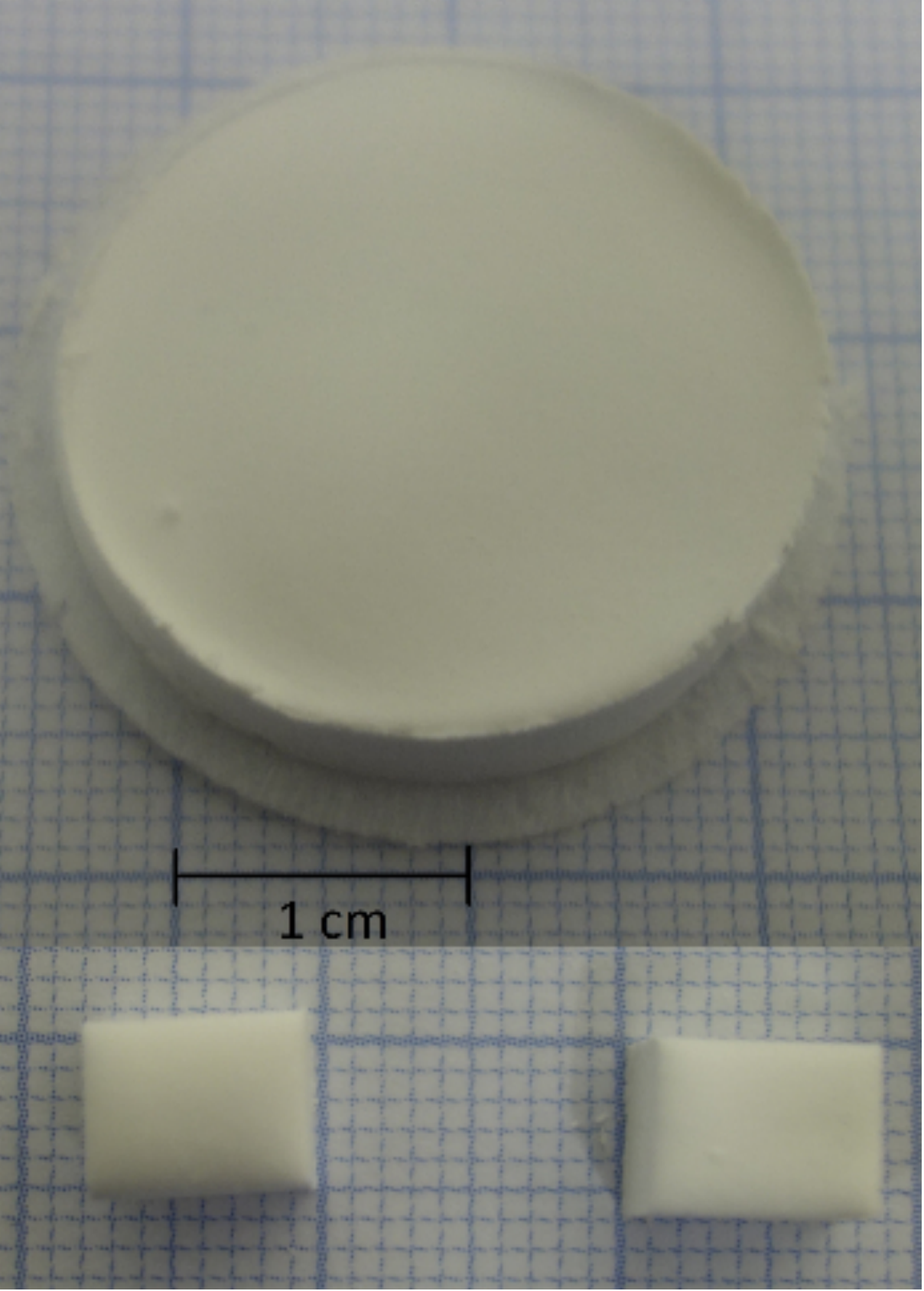}
 \caption{\label{fig:3cakes} Top: RBD dust agglomerate on its deposition filter. Bottom: cuboids cut out from a unaltered RBD dust agglomerate.}
\end{figure}

Our dust agglomerates are produced with the method described in \citet{BS2004}. A fast rotating cogwheel de-agglomerates a dust powder consisting of monodisperse 1.5 $\mu$m diameter spherical monomer grains. The monomers are coupled to a dilute gas flow and follow a siphon-shaped tube to get rid of large clumps. At the end of the tube, the dust particles are deposited on a filter on which an agglomerate with a packing density of $\Phi \approx 0.15$ is growing. As the deposition speed of the grains is much smaller than their sticking threshold, the particles form a porous body that is often referred to as random ballistic deposition (RBD). The porous agglomerate is cylindrically shaped with a diameter of 25 mm and a height of up to 7 mm. The only surface that is not altered by the contact with the filter or the walls of the tube is the top surface of the agglomerate. The samples for the mutual collision experiments (see Sect. \ref{sect:EXP}) are prepared by cutting the RBD agglomerate into cuboids such that the lateral length approximately equals the RBD agglomerate thickness (See Figure \ref{fig:3cakes}).

\subsubsection{Water-ice aggregates}\label{sect:RBDice}
\begin{figure}
  \center
  \includegraphics[width=1.\textwidth]{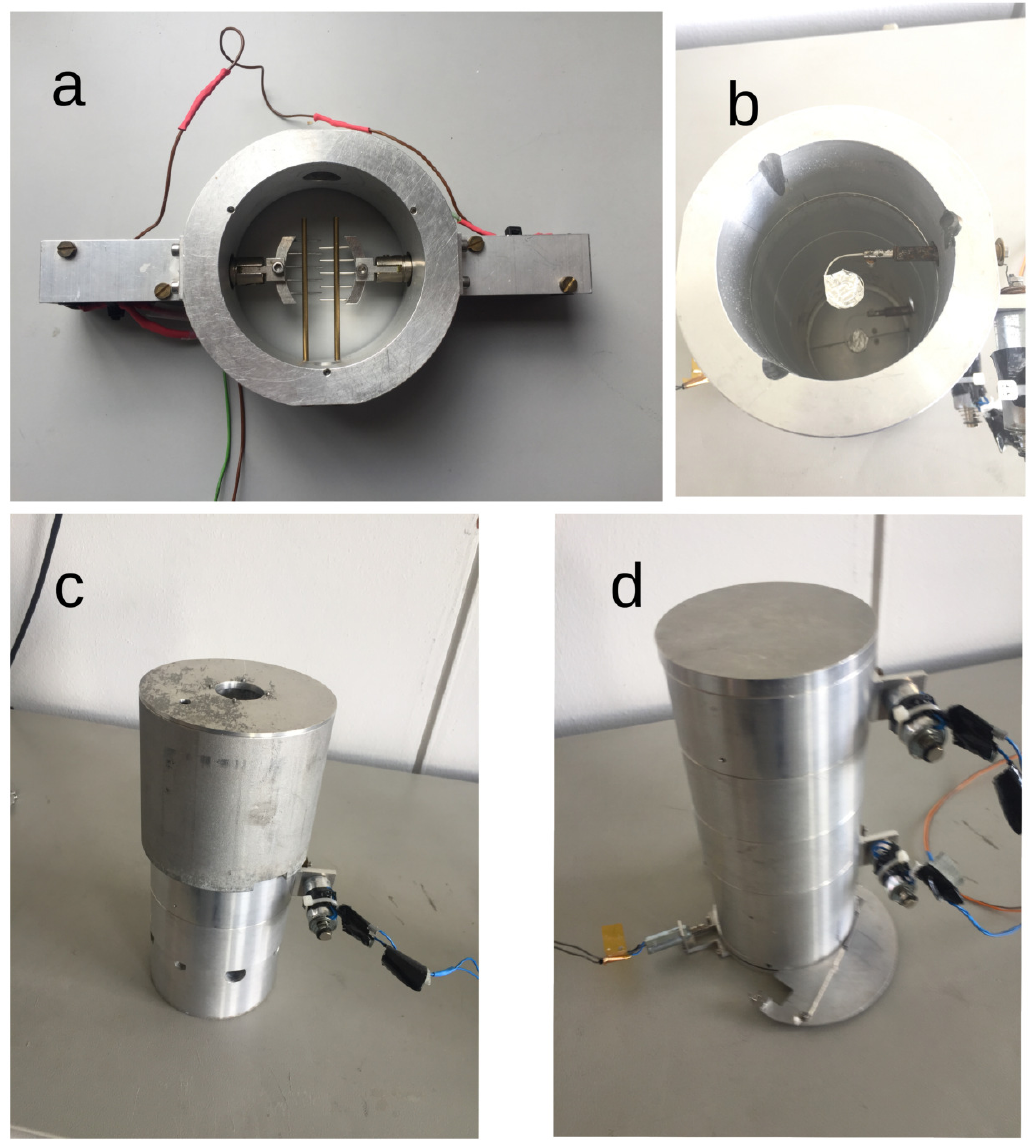}
  \caption{\label{fig:ReleaseMechanism} Experimental setup for the formation of RBD ice aggregates. (a) Type 1 water-ice aggregate release mechanism. (b) Type 2 water-ice aggregate release mechanism. (c) Cooling and growth column that consists of the cooling tube on top of the sample-release mechanism; at the bottom, a cylinder is shown whose bottom rests inside a liquid nitrogen bath and whose side holes are above the liquid level to allow the transport gas to escape. (d) The tower release unit, which consists of two type 2 release mechanisms that are arranged above each other and separated with ring spacers to increase the distance of the sample-release mechanism to enable higher relative velocities; the cylindrical tower top is closed with a cap and the bottom with a cover that can be remotely opened shortly before the drop using a solenoid (shown in open position).
}
\end{figure}

\begin{figure}
\center
    \includegraphics[width=.8\textwidth]{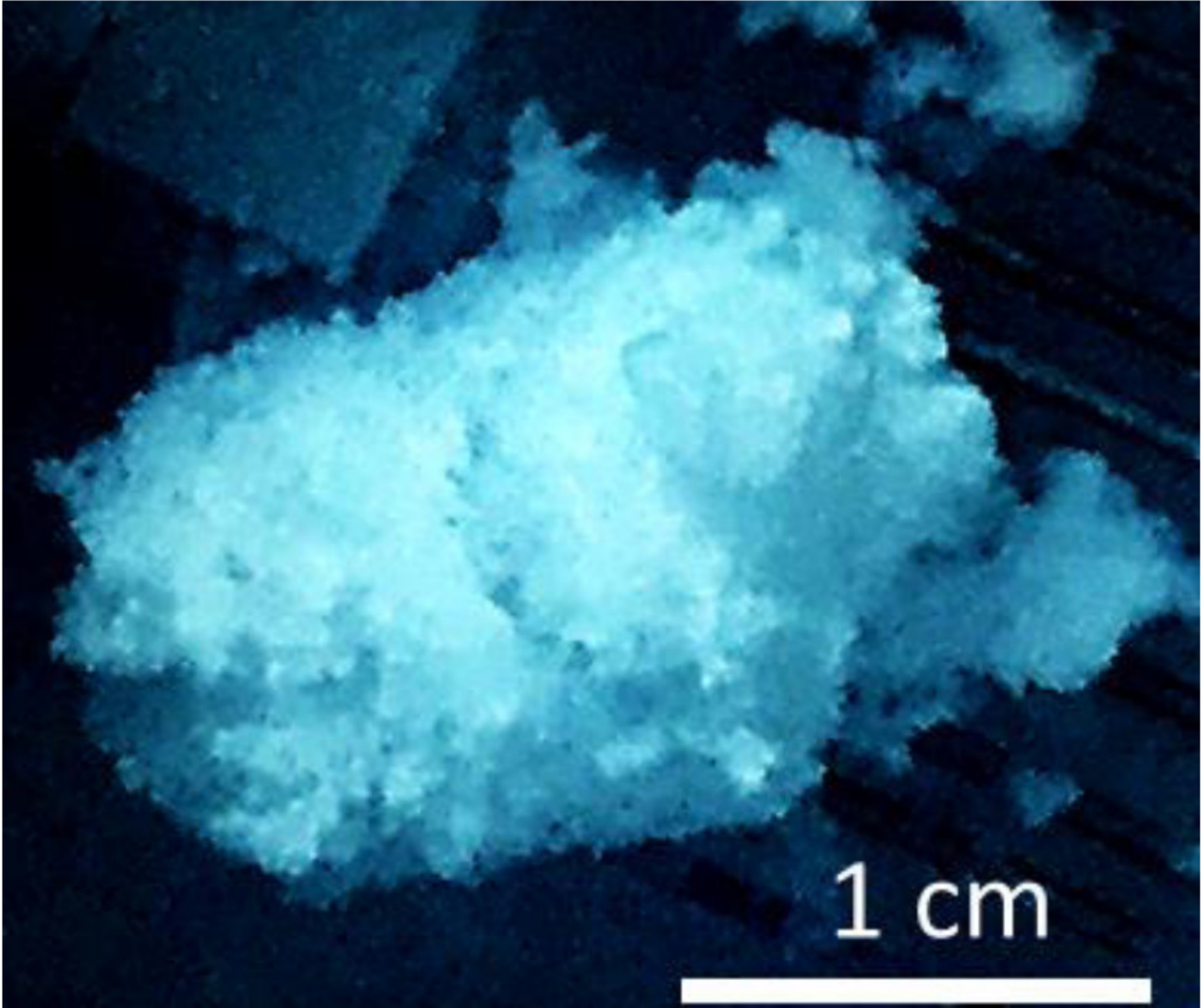}
\caption{\label{fig:IA} A water-ice agglomerate formed by the RBD process inside the growth column.}
\end{figure}

To form $\mu$m-sized spherical water-ice particles, we produced $\mu$m-sized water droplets from a water bath in which ultrasonic actuators were placed. The induced ultrasonic sound waves release water droplets with $(2.9\pm 0.8) \mu$m radius from the water bath \citep[Table 5.6]{Rezaei, Gundlachetal2011}. These droplets are coupled to a gas flow at ambient pressure and are then led through a tube that is cooled by liquid nitrogen in which the water droplets freeze almost instantaneously. In Figure \ref{fig:ReleaseMechanism} the cooling and growth column is shown. It consists of the cooling tube on top of the sample-release mechanism (see below). In the centre bore of 11 cm length, the water droplets freeze when they slowly flow though, coupled to a gas flow of about 1 cm/s speed. The annular volume around that tube is filled with liquid nitrogen. Through a hole at the bottom of the tube, the gas-ice mixture is led over the sample-release mechanism type 1 or 2 (Fig. \ref{fig:ReleaseMechanism}a and b) on which an ice agglomerate forms by the above-described RBD process (see Fig. \ref{fig:IA}). \citet{Gundlachetal2011} published long-distance-microscope images (their Figure 6) of the surface of ice agglomerates formed by the same preparation method that we used.

Below the sample-release mechanism tube, a cylinder is placed, which rests inside a liquid nitrogen bath, with its side holes above the liquid level to allow the transport gas to escape. During operation, this apparatus is inside a closed insulated styrofoam box to avoid condensation of frost.

Fig. \ref{fig:ReleaseMechanism}d shows the sample-release tower that is assembled inside a liquid-nitrogen-cooled box in which the ice agglomerates grow. Here, the type 2 release mechanisms are arranged above each other and separated with ring spacers to increase the distance of the two sample-release mechanisms to enable higher relative velocities between the two agglomerates (see Sect. \ref{sect:EXP}). The sample-release-mechanism tower is closed with a cap and a bottom lid that can be opened and placed inside the laboratory drop tower. Through the small bores, temperature sensors are mounted after the release tube is placed inside the drop tower. They monitor the temperature close to the ice agglomerates, which are at this stage passively cooled by the cylinder walls to guarantee that their temperatures are always below 130 K.  We followed this procedure also with the type 1 sample-release mechanisms.

\subsection{Properties of the RBD aggregates}
\subsubsection{Silica aggregates}
We determined the packing density $\Phi$ by weighing the cuboids (see Fig. \ref{fig:3cakes}) using a balance with 0.1 mg accuracy. The height of the cuboids was derived from the high-speed images taken during the drop-tower experiments (see Sect. \ref{sect:EXP}). The cross section of the cuboids was determined by photographing them before they were mounted on their respective release mechanisms. With known volume and mass, we determined the average packing density to be $\Phi = 0.157 \pm 0.017$. The physical properties of the silica particles and agglomerates can be found in Table \ref{table:Para2}.

\subsubsection{Water-ice aggregates}
On the sample-release mechanism, the sedimenting water-ice particles form an agglomerate with a mean packing density and standard deviation of $\Phi = 0.097 \pm 0.010$. We measured this value by cutting nine ice agglomerates with a liquid-nitrogen-cooled razor blade such that they form cuboids and measured their dimensions with a caliper. Their weight was measured with a balance that had an accuracy of 0.1 mg. To determine the packing density, we divided the ice-agglomerate mass density by the mass density of compact ice of the same temperature.

The physical properties of the water-ice particles and agglomerates can be found in Table \ref{table:Para2}.

\begin{table}
        \centering
        \caption{\label{table:Para2} Physical properties of the water-ice and silica monomers and aggregates thereof. The quantities $\Phi_1$, $\Phi_2$, $p_m$, and $\Delta$ denote the minimum and maximum filling factor at very low and very high pressures, the turnover pressure and the logarithmic width of the transition from low to high packing density, respectively.} These parameters are used to model compression curves in equation \ref{eq:gu}.
        
\begin{tabular}{l c c c}
\hline
\hline
Parameter & Water ice & Silica & References\\
\hline
Monomer radii ($\mu$m) & $2.9 \pm 0.8$ & $0.73 \pm 0.03$ & 1,2,3 \\
Monomer density &  &  &  \\ 
(kg m$^{-3}$) & 1,000 & 2,000 & \\
Surface energy &   &  &  \\
(mJ m$^{-2}$) & 20; 77  & 14 & 4,5,6 \\
Mean agglomerate & & & \\
size (m) & 0.01 & 0.006 & \\
Initial agglomerate  &  &  &\\
packing density & 0.1 & 0.15 &\\
$\Phi_1$& 0.1 & 0.15 &\\
$\Phi_2$& 0.33 & 0.33 &\\
$\Delta$& 0.33 & 0.33 &  7\\
$p_m$ (kPa)& 27 & 5.6 & 1,7 \\
\hline
References\\
1: \citet{Rezaei}\\
2: \citet{PopSchr05}\\
3: \citet{hadamcik2007}\\
4: \citet{Gundlachetal2018}\\
5: \citet{Makkonen1997}\\
6: \citet{Heim1999}\\
7: unaltered \citet{Guettleretal2009} \\
\end{tabular}
\end{table}

\section{Collisions among dust aggregates in the protoplanetary dust sub-disc}
\label{sect:CSD}
In this section, we will describe the role of mutual collisions among dust and ice aggregates after their sedimentation-induced growth. 
We use these collisions as the motivation to perform further laboratory drop-tower experiments with high-porosity dust/ice aggregates formed by the RBD process, as described in Sect. \ref{sect:RBD}. These experiments will be presented in Sect. \ref{sect:EXP}.

\subsection{Kelvin-Helmholz Instability in a dust-dominated sub-disc\label{sect:KHI}}
Once the dust/ice aggregates have settled towards the midplane and formed a dust-subdisc, collisions among equal-sized aggregates are also possible and are caused, e.g., by the Kelvin-Helmholtz instability \citep{SchrHen04,Cuzzi1993,Dubrulle}. The turbulence-driven relative velocities of \emph{compact} particles at 1 AU in such a sub-disc are 0.15 m s$^{-1}$ and 0.04 m s$^{-1}$ for particles diameters of 0.35 cm and 10 cm, respectively \citep{SchrHen04}. Because the friction time of dust particles in the Epstein regime is proportional to their density and their size, the corresponding sizes of our \emph{fluffy} dust particles are 2.3 cm  and 60 cm, respectively; for \emph{fluffy} ice particles, these sizes are 5.6 cm and 160 cm, respectively. The shear-turbulence-induced  motions are proportional to the Kepler velocity and the gas density. Because the Kepler velocity gets smaller with increasing Kepler radius, $v_\mathrm{K} \propto R_K^{-\frac12}$, relative velocities decrease with increasing Kepler radius. Moreover, the gas density in the PPD is getting smaller with with increasing distance according to $\rho_\mathrm{G} \propto R_K^{-2.75}$. Because the friction time is inversely proportional to the gas density, this also decreases the particle relative velocities caused by shear turbulence. From this simple estimation, direct growth in mutual collisions between aggregates might be possible in the dust sub-disc. In a dust-dominated sub-disc, drift motions do not occur, because the dust sub-disc is dust dominated and, thus, the gas is dragged along with Keplerian velocity. Photophoresis and photoevaporation are also not relevant, because dust-dominated discs are opaque.

\subsection{The Streaming Instability}
In case a simulation of a PPD allows a non uniform density of the dust component as well as a back-reaction of dust particles to the gas, local density enhancements form in a complex process the so called streaming instability \citep{YoudinGoodman2005}. This is an alternative way to form planetesimals. The process works best in case the particles fulfill the range $\mathrm{St} \approx 0.01 - 1$ \citep{liyoudin2021}, with the Stokes number $\mathrm{St} = \tau_f \times \Omega$ and the Kepler frequency $\Omega$. The particle sizes of our sedimentation process described in Sect. \ref{sect:SEDI} fit this requirement very well at all orbital distances and therefore enable the particle growth by the streaming instability.

\section{Experiments on low-velocity collisions among RBD agglomerates}\label{sect:EXP}
As described in the previous section, collisions among macroscopic fluffy dust agglomerates may be common in PPDs. The experiments described in Sect. \ref{sect:RBD} allowed us to produce cuboidal-shaped high-porosity dust and ice aggregates with which we performed low-velocity collision experiments.

\subsection{Sample preparation for drop-tower experiments}

\subsubsection{Silica}\label{sect:spsil}
As mentioned in Sect. \ref{sect:cuboids}, the RBD silica agglomerates were cut in cuboidal shapes for making them available to collision experiments. The cutting, however, leads to an alteration of the surface morphology with an a priori unknown coordination number of the monomers at the manipulated surface. The constituent grains of an unaltered RBD agglomerate possess coordination numbers of 2 for the interior and 1 for the surface. Whenever the surface of an RBD agglomerate comes in contact to a wall or a cutting blade, the outermost particles start to roll and increase their coordination numbers, which causes a compression and hardening of the surface. Thus, these surfaces do not represent a realistic proxy of protoplanetary aggregates grown during the sedimentation phase of the PPD dust. To avoid this, we took into account only those impacts for which at least one collision partner collided with an unaltered surface with coordination number 1. However, the other collisions were also analysed for comparison. 

\subsubsection{Water ice}\label{sect:spice}
To simulate the RBD growth of large ice aggregates, frozen water droplets were simply deposited on two cooled release mechanism, which were placed inside the sedimentation tube described in Sect. \ref{sect:RBDice}. Thus, the formed ice agglomerates are  throughout their volume with a coordination number of 2 in the interior and 1 at the free surface. 

\subsection{Drop-tower technology}\label{sect:DTT}
To perform collision experiments in the velocity range from 0.01 m s$^{-1}$ to 1 m s$^{-1}$, short-duration micro-gravity conditions are required. Our laboratory drop towers \citep[see][for details]{BBB2014} can provide up to 0.5 seconds of weightlessness, which is sufficient for our experiments. In this study, we used two evacuated laboratory drop towers, one for the high-porosity dust agglomerates and one for the high-porosity water-ice agglomerates, which will be described in some detail in the following subsections. The basic components of both drop towers are the sample-release mechanism and two free-falling high-speed cameras for the three-dimensional recording of the collision events. The sample-release mechanism is capable of releasing the upper agglomerate to free-fall $\Delta t$ earlier than the lower particle. This causes a relative speed of $v_r=\Delta t \, g$ between the two agglomerates, where $g \approx 10 \, \mathrm{m~s^{-2}}$ is the gravitational acceleration of the Earth. The distance of the two release mechanisms is adjusted such that the particles collide approximately at half their free fall time, which provides sufficient time to measure the collision speed before the impact and its outcome thereafter. The high-speed cameras are held by two solenoids and are released to free-fall at approximately $\Delta t/2$ after the release of the upper particle so that they fall approximately in the centre-of-mass frame of the two agglomerates. The agglomerates are background-illuminated using two LED arrays, which are placed opposite to the cameras outside the drop-tower glass tube. The two free-falling cameras are separated by an angle of 90 degrees to allow to retrieve full three-dimensional information about the collisions.

\subsubsection{Silica}\label{sect:DTTdust}
\begin{figure}
  \center
      \includegraphics [width=0.8\textwidth]{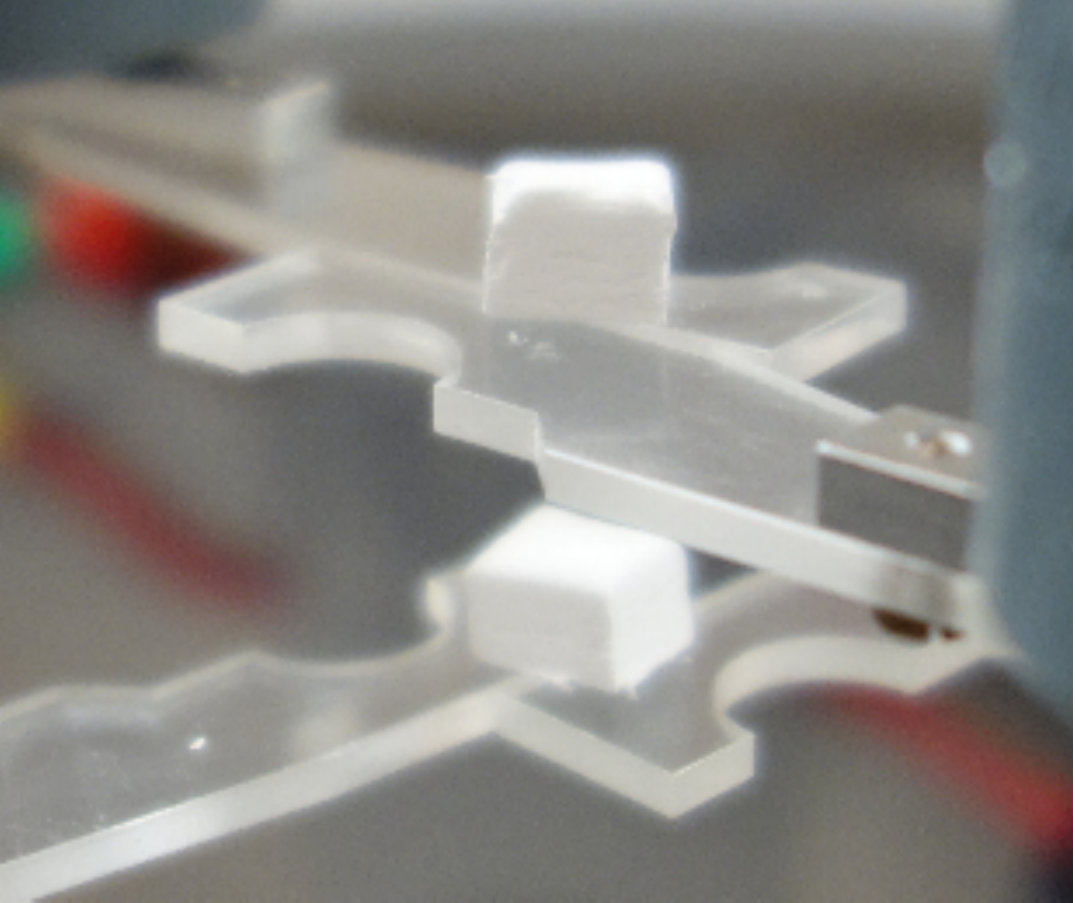}
 \caption{\label{fig:Relm} Two sample-release mechanisms with dust cuboids mounted above each other. The dust agglomerates are released when the two arms are separated abruptly. }
\end{figure}

\begin{figure*}
  \center
      \includegraphics[width=1.\textwidth]{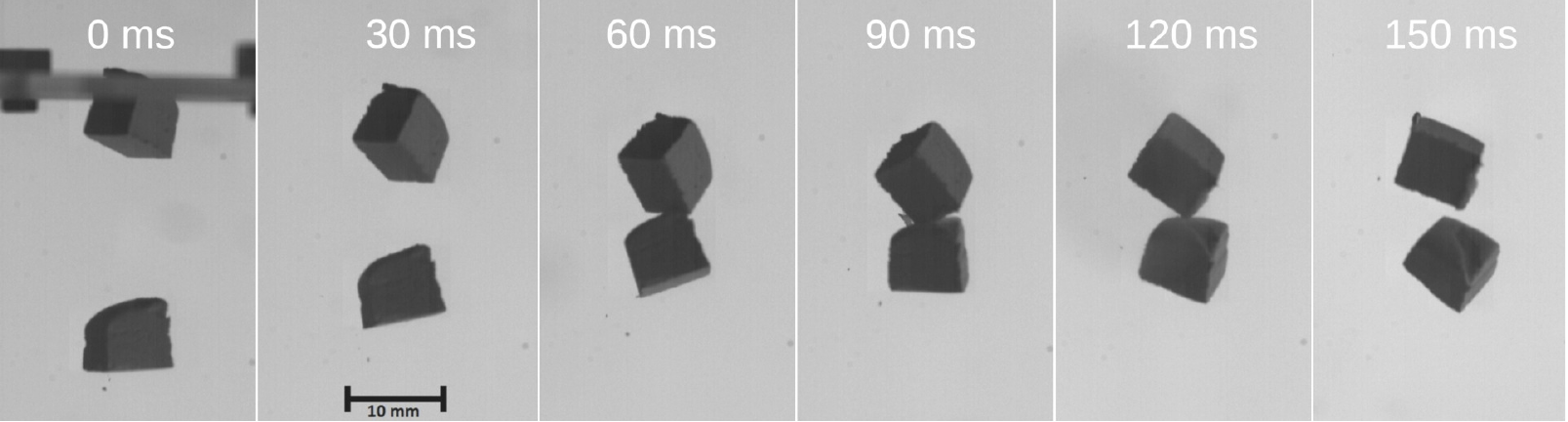}
 \caption{\label{fig:StAV} A collision of two high-porosity dust agglomerates with a reduced mass of 23 mg and a mass ratio of 0.93 at a velocity of 0.23 m s$^{-1}$. After the collision, the two dust cuboids have a relative velocity of 0.023 m s$^{-1}$, which implies a coefficient of restitution of 0.1. However, the square root of the ratio of the kinetic energies after and before the collision, including rotation, is 0.3. In the images recorded at 120 ms and 150 ms, the plastic deformation caused by the impact is visible on the surface of the lower dust agglomerate.}
\end{figure*}

To induce a collision between the dust agglomerates, each of the high-porosity silica cuboids (see Sect. \ref{sect:cuboids}) is placed on a sample-release mechanism (see Figure \ref{fig:Relm}). The two sample-release mechanisms themselves are placed above each other in the upper part of the laboratory drop tower, as described by \citet{Beitz2011}. The drop tower consists of a glass tube of 1.5 m length and an inner diameter of 0.22 m. Two vacuum chambers are placed on the top and at the bottom of the vacuum tube, respectively. Inside the upper vacuum chamber, the sample-release mechanisms are placed. Each of the two sample-release mechanisms consist of two arms that are mounted on opposite sides. The dust agglomerates are placed on both arms simultaneously and can be released to free-fall when the arms are oped abruptly by the use of solenoids. Releasing the two dust cuboids with a given time difference, the relative velocity between the agglomerates can be adjusted. Figure \ref{fig:Relm} shows an example of the placement of the two dust cuboids and their corresponding sample-release mechanisms. An example of a high-speed image sequence from a dust-agglomerate collision is shown in Figure \ref{fig:StAV}.

\subsubsection{Water ice}\label{sect:DTTice}
The cryogenic drop tower consists of a glass tube of 1.4 m length and 0.3 m inner diameter. Two vacuum chambers are connected on both ends to the glass tube. Inside the upper vacuum chamber, the sample-release mechanisms are placed. 

We used two types of sample-release mechanisms for ice aggregates, hereafter called type 1 and type 2, respectively. The type 1 sample-release mechanism (see Figure \ref{fig:ReleaseMechanism}a) consists of two inter-meshing forks. To release the ice agglomerate, the forks are rapidly pulled apart. This is done with both ice agglomerates, which are placed with their respective release mechanisms above each other. The upper ice agglomerate is released earlier and therefore gains relative velocity to the lower and later-released ice agglomerate. A side effect of the type 1 mechanism is that the agglomerates are torn apart, resulting in a release of agglomerate fragments. The feature of this behaviour is that we can also investigate collisions of small agglomerates with a velocity dispersion, including very small velocities from collisions of fragments from the same agglomerate. Larger relative motions arise from collisions of fragments stemming from different sample-release mechanisms. 

The type 2 sample-release mechanism (see Figure \ref{fig:ReleaseMechanism}b) consists of a surface on which the ice-agglomerate grows. Two release mechanisms were placed on top of each other in the drop tower. The arms of the release mechanisms rotate downwards with a high acceleration to release the agglomerates by inertia into free fall. As above, the upper ice agglomerate is released earlier than the lower one to gain relative speed. The time difference of the two releases and the distance of the sample-release mechanism are adjusted such that the agglomerates collide typically after about 0.25 s, which is half their free-fall time. To avoid sintering of the water-ice agglomerates, we continuously controlled their temperature and made sure that the temperature was always below 130 K. \citet{Gundlachetal2018} computed an analytical sintering model for ice agglomerates and show in their Figure 12 the sinterneck-formation timescale as a function of temperature. They found that at 130 K, a sinterneck starts to form after 28 hours. As our maximal experiment duration, beginning with the formation of the RBD water-ice agglomerates and ending with their collision in the drop tower, is about one hour, our measurements are unaffected by the sintering process. Moreover, most of the time our samples are stored inside an aluminium cylinder, which is placed in a nitrogen bath at 77 K at which the sinterneck-formation timescale is almost infinite. During the experiment preparation time, the water-ice agglomerates are stored between 77 K and 130 K for only eight minutes.

\begin{figure*}
  \center
      \includegraphics[width=.9\textwidth]{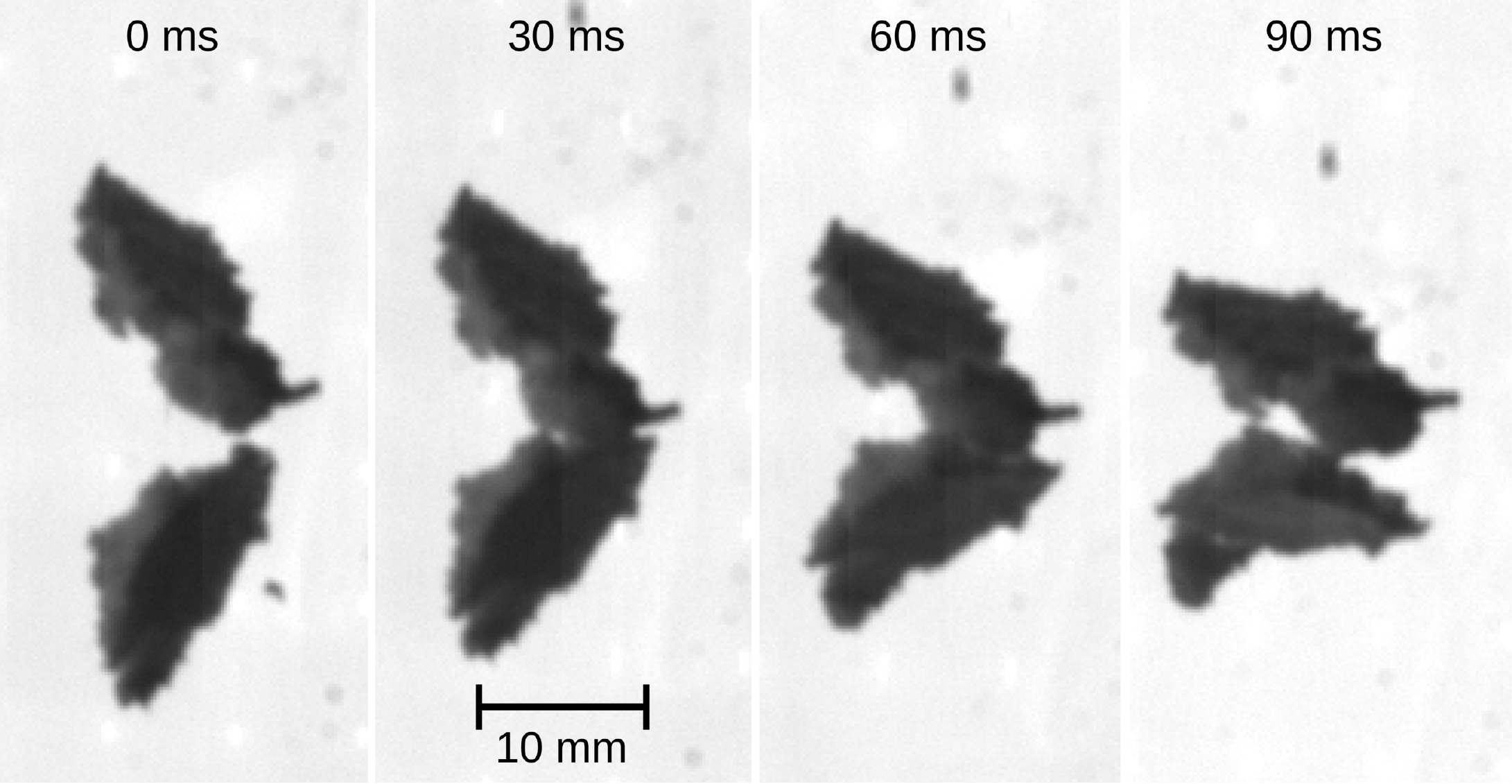}
 \caption{\label{fig:Impact} Collision of two water-ice agglomerates with a reduced mass of 29 mg and a mass ratio of 0.93 at a collision speed of 0.25 m s$^{-1}$. After the impact, the two agglomerates possess a relative velocity of 0.017 m s$^{-1}$, which implies a coefficient of restitution of 0.07.}
\end{figure*}

An example of a high-speed image sequence from a collision between two water-ice agglomerates in our drop tower is shown in Figure \ref{fig:Impact}.

\subsection{Results of the collision experiments}
With the technology described in Sect. \ref{sect:DTT}, we performed low-velocity collision experiments with dust (see Sect. \ref{sect:spsil}) and ice aggregates (see Sect. \ref{sect:spice}), the results of which we present in the following subsections.

\subsubsection{Silica}\label{sect:ResSilica}
\begin{figure}
  \center
      \includegraphics[width=0.6\textwidth]{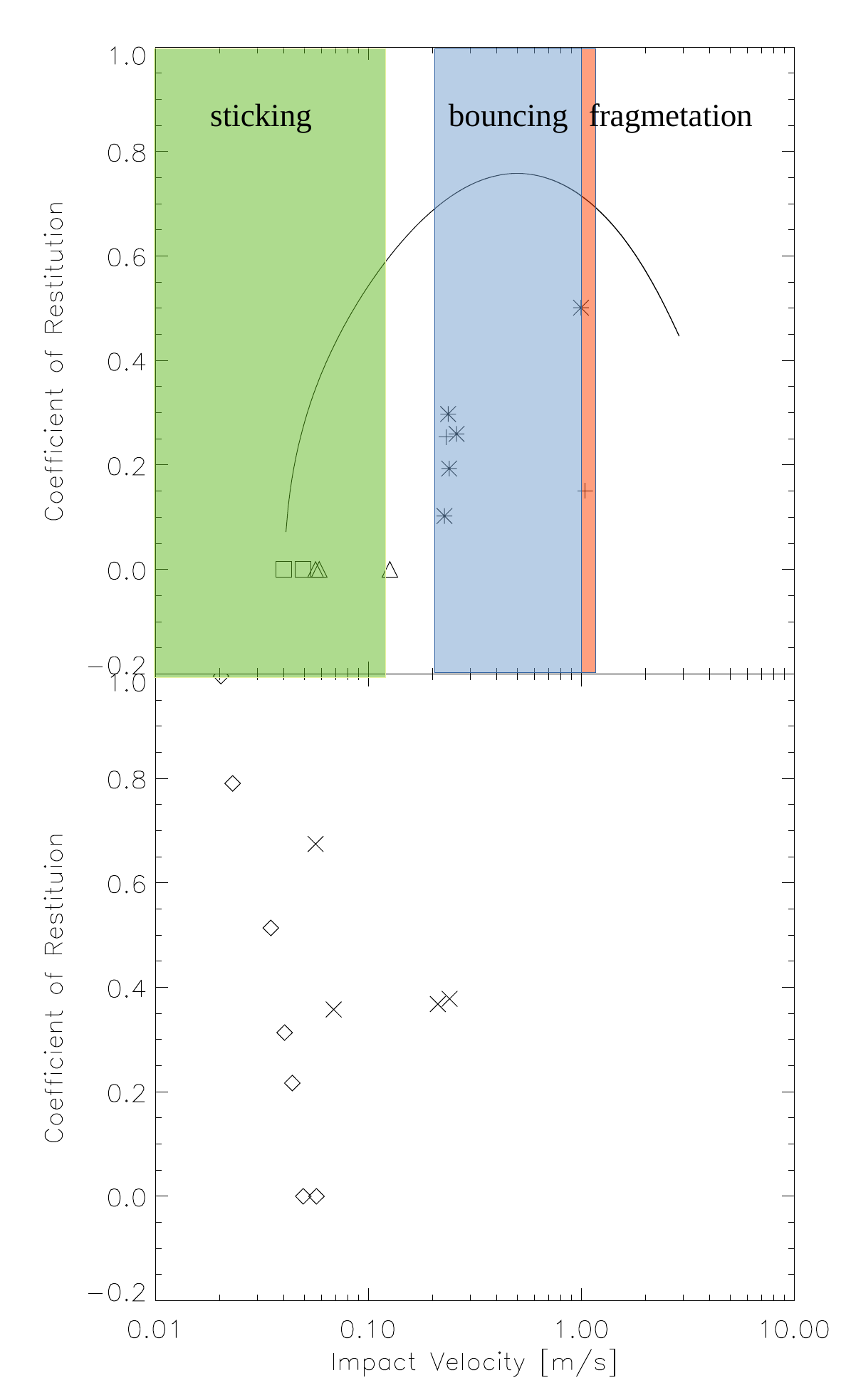}
 \caption{\label{fig:RCDDE} Coefficient of restitution for high-porosity silica agglomerates as a function of impact velocity. Top: Experiments for which at least one of the agglomerates collided with an unaltered surface. Asterisks denote collisions of an unaltered surface with an edge, crosses mark collisions of an unaltered surface with a processed/compacted surface, triangles stand for collisions of two unknown surfaces, squares stand for collisions of an unknown surface with an edge. Unknown surfaces could not be determined because of poor viewing angles of the cameras, but might likely be unprocessed. From five experiments at $\sim 1 \, \mathrm{m ~ s^{-1}}$ collision speed, three particle pairs totally fragmented and are not shown here, one experiment led to bouncing (at a coefficient of restitution of 0.5) and one led to bouncing with mass transfer (at a coefficient of restitution 0.15). The solid line is our impact model using the monomer surface energy of $\gamma$=14 mJ/m$^{-2}$ from \citet{Heim1999}. Bottom: Experiments for which both agglomerates collided with their processed/compressed surfaces. Crosses denote collisions of a compressed surface with an edge, diamonds mark collisions of an edge with an edge. The color coding shows the regions where sticking bouncing and fragmentation was found by our experiments.
 }
\end{figure}

The cut silica cuboids (see Sect. \ref{sect:spsil}), which we used in the drop-tower experiments, possessed masses in the range of $1.6 \times 10^{-5}$ kg to $1.3 \times 10^{-4}$ kg. The collision speeds ranged from $\sim 0.02 \, \mathrm{m ~ s^{-1}}$ to $\sim 1 \, \mathrm{m ~ s^{-1}}$. We determined the translational and rotational relative speeds of the colliding particles before and after the collision and determined the coefficient of restitution, including the rotational motion of the agglomerates, using the method of \citet{SBSK} (their Eqs. 1 and 2). In the upper graph of Figure \ref{fig:RCDDE}, the coefficient of restitution of the observed collisions of the silica cuboids is shown as a function of impact velocity for those collisions in which at least one unaltered surface participated in the collisions (see Sect. \ref{sect:spsil}), where a coefficient of restitution of zero stands for sticking and values above zero denote bouncing (if not otherwise noted). The asterisks mark collisions of an unaltered surface with an agglomerate edge, pluses mark collisions of an unaltered surface with a processed surface, triangles mark collisions of two surfaces with unknown status, and squares denote collisions of an unknown surface type with the edge of an agglomerate. Please note that it was not possible to explore the surface status of particles that had a collision velocity below $0.1 \, \mathrm{m ~ s^{-1}}$, because the distances of the particles in the high-speed camera images were simply too small to observe their surfaces. However, due to the preparation method, it is likely that at least the lower particle's surface was unaltered, thus these measurements are included in the upper part of Figure \ref{fig:RCDDE}. It can be seen in the data that particles stick to each other for velocities up to $0.12 \, \mathrm{m ~ s^{-1}}$. For collision velocities around $0.2-0.3 \, \mathrm{m ~ s^{-1}}$, bouncing occurs with coefficients of restitution between 0.1 and 0.3. We also observed five collisions at an impact speed of $\sim 1 \, \mathrm{m ~ s^{-1}}$. Three of these experiments led to the total disruption of both collision partners. One collision of an unaltered surface with the edge of the other dust agglomerate led to bouncing. One collision of an unaltered surface with a processed surface led to bouncing with mass transfer, which means that from one particle a small part broke off and got stuck on the second particle. These results imply that the fragmentation barrier for low packing density dust agglomerates is close to $1 \, \mathrm{m ~ s^{-1}}$. The coefficient of restitution for the two bouncing collisions at $\sim 1 \, \mathrm{m ~ s^{-1}}$ impact speed is between 0.15 and 0.5. Contrary to the predictions by \citet{wadaetal2011}, we observed an extended velocity range in which bouncing occurred between those for sticking and fragmentation.

In the lower graph of Figure \ref{fig:RCDDE}, the coefficients of restitution for collisions between dust cuboids with their processed and, thus, compacted surfaces are shown. Here, the crosses denote collisions between a compressed surface and an edge and the diamonds denote collisions between two edges. In contrast to the unaltered surfaces, the particles that collide with compacted surfaces show a decreasing coefficient of restitution with increasing impact velocity, as was found before for compressed particles \citep[e.g.][]{SBSK}. This effect is discussed in \citet{wadaetal2011}. 

We took great care in keeping the surface of our ice and dust particles unaltered. As shown above and in \citet{wadaetal2011}, the surface condition has a fundamental effect on the outcome of a collision. A possible cause for the discrepancy between our experimental findings and the model predictions by \citet{wadaetal2011} is the different internal structure of the agglomerate used in the experiments and the model. Our agglomerates are formed by the "hit and stick" mechanism, in which the monomers impact at random positions and stick upon the first contact. This leads to agglomerates that are made up of particle chains that are weakly interconnected. In our experiments, particles only impact from one direction, whereas in the protoplanetary disc the agglomerates would supposedly rotate. This would also lead to a chain-like particle structure, with the chains directed radially. On the other hand, \citet{wadaetal2011} designed their agglomerates by randomly removing particles from a densely packed structure until the desired filling factor was reached. This leads to an isotropic packing.

 The rolling threshold force used in most publications is the rolling threshold torque of two particles \citep{DominikTielens1997}, divided by the sum of their radii. However, this is only correct if the force is perpendicular to the particle alignment. In case of a chain-like structure, the impact direction is more likely along the chain alignment direction so that a much higher rolling force is required for compaction, which may lead to harder agglomerates and a reduction in their capacity to absorb kinetic energy. This, in turn, will reduce the sticking threshold velocity.

\subsubsection{Water ice}
\begin{figure}
  \center
      \includegraphics[width=0.7\textwidth]{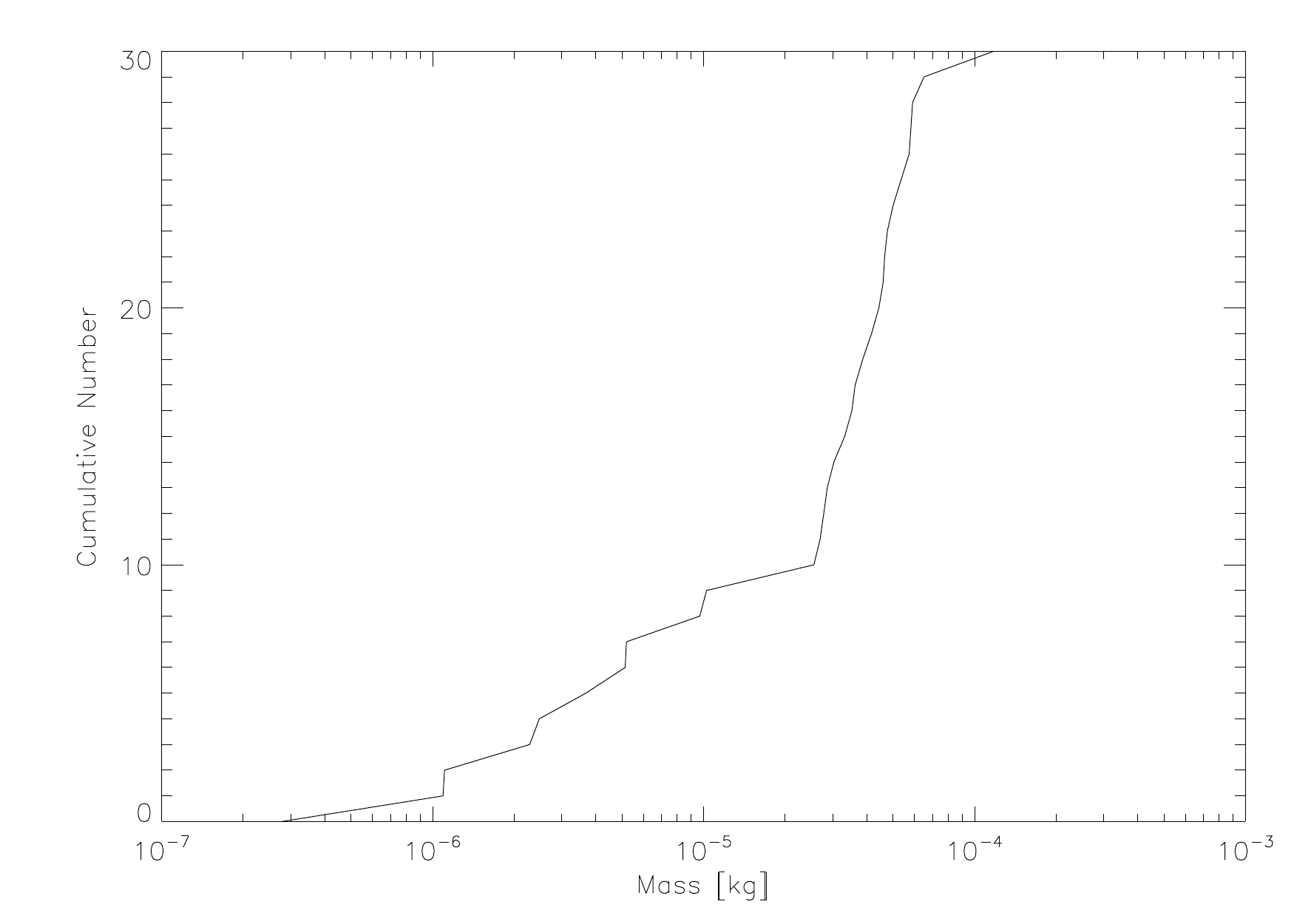}
 \caption{\label{fig:CMP} Cumulative number of collisions as a function of the reduced masses of the collisional partners in the water-ice-agglomerate collisions. The mass range from $3\times10^{-7}$ to $10^{-5}$ kg corresponds to the experiments performed with the type 1 sample-release mechanism, the mass range from $2.5\times 10^{-5}$ kg to $6.5\times 10^{-5}$ kg corresponds to the experiments with the type 2 sample-release mechanism.}
\end{figure}

The water-ice RBD agglomerates, which we used in the drop-tower experiments, possessed masses in the range of $3 \times 10^{-7}$ kg to $6.5 \times 10^{-5}$ kg. As pointed out in Sect. \ref{sect:DTTice}, this wide range of masses stems from the usage of two different sample-release mechanisms. As the type 1 sample-release mechanism led to the fragmentation of the water-ice agglomerates, the observed collisions are amongst much smaller aggregates than for the type 2 mechanism. Fig. \ref{fig:CMP} shows the cumulative mass-frequency distribution function of all water-ice agglomerates in the analysed collisions. The masses between $3 \times 10^{-7}$ kg and $1 \times 10^{-5}$ kg stem from the type 1 experiments, the masses in the range of $2.5 \times 10^{-5}$ kg and $6.5 \times 10^{-5}$ kg result from the type 2 experiments. The collision speeds ranged from $\sim 0.03 \, \mathrm{m ~ s^{-1}}$ to $\sim 0.3 \, \mathrm{m ~ s^{-1}}$ for the experiments using the type 1 mechanism and from $\sim 0.15 \, \mathrm{m ~ s^{-1}}$ to $\sim 0.62 \, \mathrm{m ~ s^{-1}}$ for the experiments using the type 2 mechanism. 

\begin{figure}
  \center
      \includegraphics[width=0.6\textwidth]{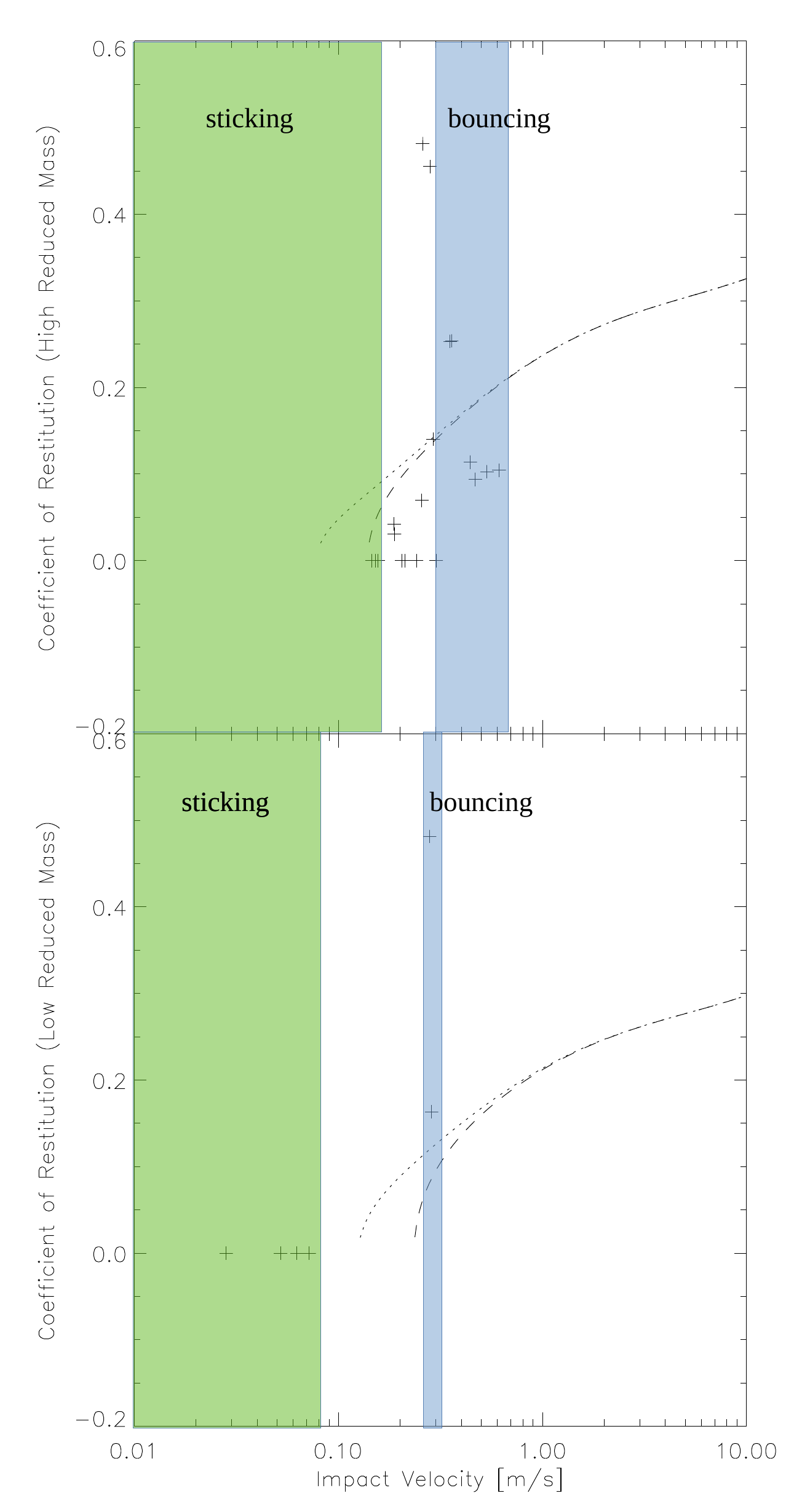}
 \caption{\label{fig:RCIDE} Coefficient of restitution for high-porosity water-ice agglomerates as a function of impact velocity. Top: Using the type 2 sample-release mechanism, the water-ice-agglomerate masses ranged from $2.5 \times 10^{-5}$ kg to $6.5 \times 10^{-5}$ kg (mean agglomerate mass of $4.14 \times 10^{-5}$ kg) and the collision velocities ranged from $\sim 0.15 \, \mathrm{m ~ s^{-1}}$ to $\sim 0.62 \, \mathrm{m ~ s^{-1}}$. Bottom: Using the type 1 sample-release mechanism, the water-ice-agglomerate masses ranged from $3 \times 10^{-7}$ kg to $1 \times 10^{-5}$ kg (mean agglomerate mass of $4.74 \times 10^{-6}$ kg) and the collision velocities ranged from $\sim 0.03 \, \mathrm{m ~ s^{-1}}$ to $\sim 0.3 \, \mathrm{m ~ s^{-1}}$.  The curves show our impact model (see Sect. \ref{sect:MODEL}) using mean reduced masses and a monomer surface energy of $\gamma$=20 mJ m$^{-2}$ \citep{Gundlachetal2018} (dotted curves) and $\gamma$=77 mJ m$^{-2}$ \citep{Makkonen1997} (dashed curves), respectively. The color coding shows the regions where sticking and bouncing was found by our experiments.}
\end{figure}

From the high-speed images recorded at 1,000 fps during the free-fall of the water-ice aggregates, we analysed, using both camera views, the relative translational and rotational motion of the colliding particles before and after the impact. The coefficient of restitution was calculated using the method of \citet{SBSK} (their Eqs. 1 and 2), taking the rotational motion into account. However, we found that the influence of the ice-particle rotation on the coefficient of restitution was negligible. In Figure \ref{fig:RCIDE}, the coefficient of restitution is shown for experiments with the type 2 sample-release mechanism (top graph) and with the type 1 sample-release mechanism (bottom graph), where a coefficient of restitution of zero stands for sticking and values above zero denote bouncing. 

The steep peak of the coefficient of restitution close to the sticking threshold velocity is due to the so-called Bingham behaviour. This is a non-Newtonian effect caused by a friction threshold below which relative motion between monomers are not possible \citep[see e.g.][]{Cioru2016,Betten2013}. In our case, this is the rolling threshold force of the monomers. The force has to be generated by the momentum change $\frac{dp}{dt}$ of the impacting particles.  As long as this threshold is not exceeded, the impact energy cannot be damped efficiently by rolling and sliding of the monomers in the agglomerates so that the agglomerate has the properties of a rigid body and therefore a higher coefficient of restitution. For a better understanding of this effect, we performed an exemplary calculation in Appendix \ref{APP:BH} of how the Bingham effect may matter for our agglomerates. The Bingham behaviour is not modelled in the calculation of Section \ref{sect:MODEL}.

From the data shown in Figure \ref{fig:RCIDE}, it is evident that also for the high-porosity water-ice agglomerates, which possess coordination numbers of not more than 2, a transition from sticking to bouncing occurs, in contrast to the model by \citet{wadaetal2011}. The sticking threshold for the water-ice agglomerates in both cases falls between $\sim 0.1 \, \mathrm{m ~ s^{-1}}$ and $\sim 0.3 \, \mathrm{m ~ s^{-1}}$, but cannot be constrained better than that. In order to understand this sticking threshold and to derive its mass dependency, we developed a collision model for RBD agglomerates, which will be described in the next section.

\section{Modelling the collisions between high-porosity agglomerates}\label{sect:MODEL}
The challenge of building a model for collisions between porous media is that their packing density changes during impact. This means that their fundamental properties (e.g Young's modulus, Poisson number, viscosity) are not constant, but dependent on impact velocity and time.

To model our findings described above, we used the theory of viscoelastic spheres \citep{Brilliantov2005}. However, it is not feasible to integrate into that model parameters that change during the collisional process. Instead, we used the impact-velocity-dependent parameters at the final compression. The corresponding final packing density of the agglomerates is calculated from their impact velocity, considering the kinetic and compression energy. This again is only a zero-order estimation. This assumption works in an acceptable way for the Young's modulus, quite well for the Poisson ratio and poorly for the viscosity (see the discussion in the following sections). However, apart from the Bingham behaviour, the major part of the collision dynamics happens in the high-compression state towards the end of the intrusion phase where the agglomerate is compressed close to its maximum value. Moreover, the Bingham behaviour is limited to a very small velocity range. Therefore it only plays a minor role in the overall coagulation process. The simplifications adopted here, especially for the viscosity, might be the reason that our model does not resolve the measured Bingham behaviour, which is beyond the scope of this paper. The surface energy of the dust-sample material of 14 mJ m$^{-1}$ is taken from \citet{Heim1999}. To test the outcomes of our model for the water-ice agglomerates, we used the surface energies found by \citet{Makkonen1997} (77 mJ m$^{-1}$) and \citet{Gundlachetal2018} (20 mJ m$^{-1}$). Details of the model can be found in Appendix \ref{APP:CM}.

\subsection{Application of the dust-aggregate collision model: the sticking threshold as a function of agglomerate size}
\begin{figure}
  \center
      \includegraphics[width=0.8\textwidth]{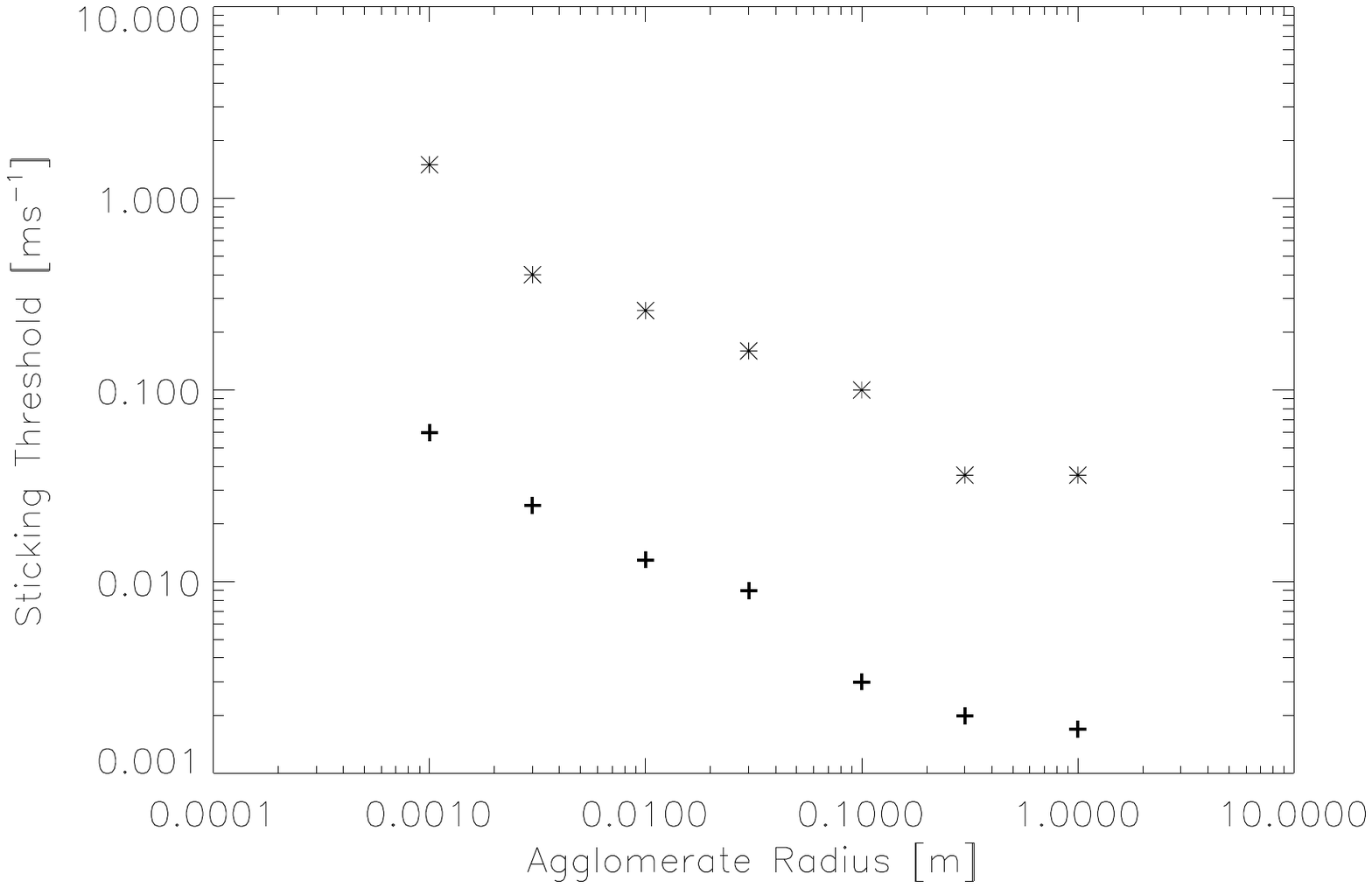}
 \caption{\label{fig:StSiz} Model results of the sticking threshold of water-ice agglomerates with a packing density of $\phi=0.1$ and a surface energy of 77 mJ m$^{-1}$ (asterisks) as well as of dust agglomerates with a packing density of $\phi=0.15$ and a surface energy of 14 mJ m$^{-1}$ (pluses) for various agglomerate radii. 
  }
\end{figure}

Based on the model by \citet{Brilliantov2005}, we calculated the sticking threshold for porous water-ice agglomerates with a packing density of $\phi=0.1$ and a surface energy of 77 mJ m$^{-1}$ \citep{Makkonen1997} as well as for dust agglomerates with a packing density of  $\phi=0.15$ for different agglomerate radii. The results are shown in Figure \ref{fig:StSiz}. As expected, the sticking threshold velocity decreases with increasing agglomerate size. Although the surface energy of the water-ice agglomerates is only a factor $\sim 5$ larger, their threshold velocity for sticking exceeds that of the dust agglomerates by more than an order of magnitude. This finding is in agreement with the experimental results by \citet{Gundlach2015}.

\subsection{Application of the dust-aggregate collision model: the sticking threshold as a function of monomer size}
\begin{figure}
  \center
      \includegraphics[width=0.8\textwidth]{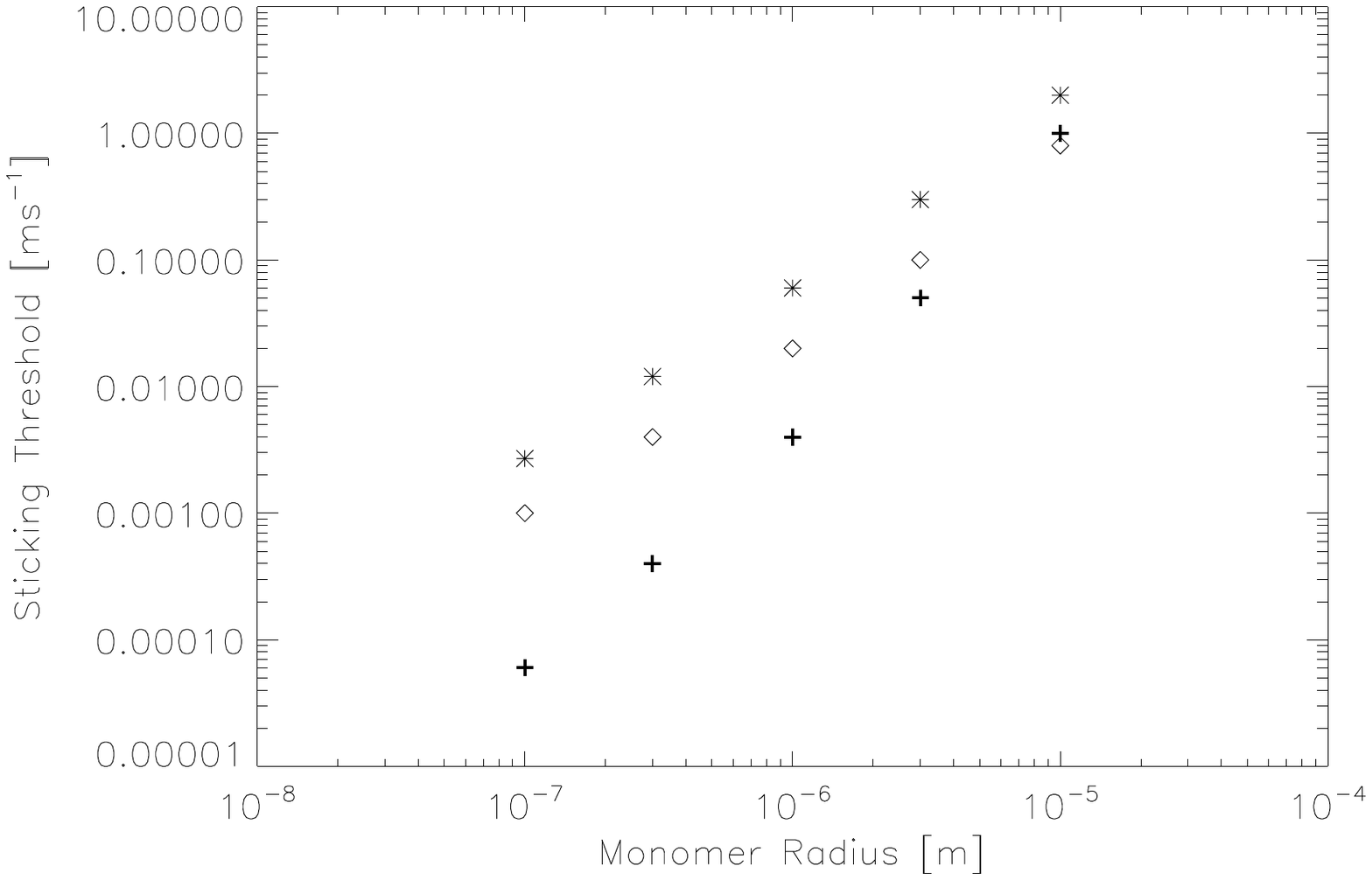}
 \caption{\label{fig:StMon} Model results of the sticking threshold of water-ice agglomerates with 3 cm radius, a packing density of $\phi=0.1$ and a surface energy of 77 mJ m$^{-1}$ (asterisks) and 20 mJ m$^{-1}$ (diamonds) as well as of dust agglomerates with 3 cm radius, a packing density of $\phi=0.15$ and a surface energy of 14 mJ m$^{-1}$ (pluses) for various monomer-grain radii. 
  }
  
\end{figure}

In Appendix \ref{APP:CM}, we derive the physical properties of dust agglomerates from their experimentally derived quasi-static compression curves (Eq. \ref{eq:gu}) for a given monomer-particle size. In our past work \citep[see][]{SBBG}, we found that the  monomer radius $r_m$ of an agglomerate only affects the parameter $p_m$ in Eq. \ref{eq:gu} and gave an analytical expression for its dependency on the monomer-grain size \citep[see Eq. 9 or solid line in Figure 6 of][]{SBBG}. Based on the model by \citet{Brilliantov2005}, we calculated the sticking threshold for porous water-ice agglomerates with a packing density of $\phi=0.1$ and a surface energy of 77 mJ m$^{-1}$ \citep{Makkonen1997} and 20 mJ m$^{-1}$ \citep{Gundlachetal2018} as well as for dust agglomerates with a packing density of  $\phi=0.15$ for different monomer-grain radii and show the results in Fig. \ref{fig:StMon}. 

\citet{DominikTielens1997} found the rolling torque $M_r$ to be proportional to $r_m$. Therefore the rolling force $F_r = M_r /  r_m$ and the rolling threshold force are independent of $r_m$. The number of monomers at the contact surface of an agglomerates during impact is proportional to $r_m^{-2}$. Thus, the deformation pressure and the elastic threshold of an agglomerate are proportional to $r_m^{-2}$. In other words, smaller monomer particles have the same rolling force, but the impact pressure is distributed to many more particles. For the forces of pull-off ($\propto r_m$), sliding ($\propto r_m^{\frac43}$) and twisting ($\propto r_m$) \citep{DominikTielens1997}, this effect is weaker but also present.

A stunning feature is that the sticking threshold velocity increases with monomer size (see Fig. \ref{fig:StMon}). As discussed above, agglomerates will get "softer" in case they consist of larger monomers, because rolling, sliding, pull-off and twisting forces increase slower with increasing monomer-grain size than the number of monomers decreases at the agglomerate-agglomerate contact area. This causes a larger deformation and enables more energy dissipation for larger monomer sizes. As can also be seein in Fig. \ref{fig:StMon}, an increase in surface energy of the monomers leads to an increase of the sticking threshold velocity, because the adhesion energy is increased. When the packing density is increased, the coordination number is also increased, which leads to a lower sticking threshold velocity and a steeper dependency on the monomer-grain size. The latter might be caused by a different force-chain distribution inside the agglomerate. For increasing monomer sizes, the fragmentation threshold velocity decreases \citep{DominikTielens1997} so that the bouncing regime, as an intermediate range between sticking and fragmentation, may only be found for monomer-grain sizes below a threshold value.

\section{Conclusions}
We presented a simple sedimentation and growth model for dust and ice particles in protoplanetary discs and calculated the maximum size of the agglomerates formed during sedimentation as a function of distance to the central star. We experimentally produced realistic high-porosity dust and ice agglomerates as analogues to the sedimentation-grown particles in PPDs by simulating the sedimentation and subsequent growth from micrometre-sized grains to cm-sized agglomerates with different methods for dust and ice. To assist simulations of dust growth by the streaming instability or shear turbulence in the midplane of a protoplanetary disc \citep[e.g.][]{YoudinGoodman2005,SchrHen04}, we also experimentally simulated collisions among  macroscopic high-porosity dust and ice agglomerates, using our laboratory drop tower \citep{BBB2014}, and measured the coefficient of restitution and the sticking threshold. We found a qualitatively similar behaviour for our dust and ice agglomerates with sticking thresholds of $\sim 0.1$ ms$^{-1}$ and coefficients of restitution for velocities larger than $\sim 0.3$ ms$^{-1}$ of $\sim 0.1$ and $\sim 0.2$ for ice and dust agglomerates, respectively. The low packing density of our ice ($\Phi=0.1$) and dust agglomerates ($\Phi=0.15$) corresponds to a mean coordination number of the monomer particles of $\lesssim 2$ \citep{Langemaat}. Theoretical considerations of \citet{wadaetal2011} imply that bouncing should not be occurring at low coordination numbers. However, our experiments show that bouncing happens in a velocity range between the sticking and the fragmentation thresholds.
We took great care in keeping the surface of our ice and dust agglomerates unaltered. As shown by our experiments and in \citet{wadaetal2011}, the surface condition has a fundamental effect on the outcome of a collision. A possible cause for the discrepancy between our experimental results and those of the model by \citet{wadaetal2011} may be the different internal structures of the used agglomerates (see Sect. \ref{sect:ResSilica}).

Finally, we developed a method to gain the fundamental properties of the dust and ice agglomerates used in our study, like Young's modulus, Poisson ratio, shear viscosity and bulk viscosity. With these parameters, it was possible to estimate the coefficient of restitution, using the model of \citet{Brilliantov2005}, which fits our measurements.

This work shows that agglomerates with low packing densities of $\Phi=0.1-0.15$ stick to one another for relative velocities up to about 10 centimetre per second and that bouncing occurs for velocities above this threshold. 

\section{Summary}
\begin{enumerate}
  \item We produced cm-sized agglomerates with RBD structures, consisting of $\mathrm{\mu m}$-sized $\mathrm{SiO_2}$ grains at room temperature and water-ice particles at temperatures between 77 K and 130 K, as anlogues to the agglomerates grown during the dust-sedimentation phase of PPDs
  \item We performed low-velocity ($\lesssim 1~\mathrm{m~s^{-1}}$) collisions between these RBD agglomerates and found sticking for velocities $\lesssim 0.1~\mathrm{m~s^{-1}}$ and bouncing for velocities between $\sim 0.1~\mathrm{m~s^{-1}}$ and $\sim 1~\mathrm{m~s^{-1}}$.
  
  \item Bouncing occurs also for aggregates with coordination numbers $\lesssim 2$, in contrast to predictions by numerical models \citep{wadaetal2011}. It should, however, be noted that the internal structures of the experimental and numerical aggregates were different, although the coordination numbers were both low. 
  \item We developed a numerical model that describes the collision behaviour (sticking threshold, coefficient of restitution in the bouncing regime) of the experimental aggregates.
  \item This model predicts that ice aggregates are much stickier than dust aggregates of the same size and that the threshold velocity for sticking decreases with increasing aggregate size (see Figure \ref{fig:StSiz}).
\end{enumerate}

{\bf Acknowledgments:}
This research was supported by the Deutsches Zentrum f\"ur Luft- und Raumfahrt under grants no. 50WM1236, 50WM1536, 50WM1846 and the Deutsche Forschungsgemeinschaft under grant no. BL298/24-1. The Authors thank Coskun Aktas for his help in improving our experimental setup.

\vskip0.3cm
{\bf Data availability} \\
All data are incorporated into the article.

\bibliographystyle{unsrtnat}
\bibliography{ms}

\newpage
\vskip5.0cm


\appendix

\section{\label{APP:CM}A collision model for high-porosity dust aggregates}

\subsection{Young's Modulus}\label{sect:YM}

The Young's modulus of a granular medium is dependent on its packing density. We start from the packing density $\Phi$ as a function of compression  pressure $\Sigma$ from \citet{Guettleretal2009,SBBG} (in the following, lg and ln mean the logarithm to the base $10$ and the natural logarithm, respectively).
\begin{equation}
    \Phi(\Sigma)=\Phi_2 -\frac{\Phi_2 - \Phi_1}{\exp\left(\frac{\lg \Sigma -\lg p_m}{\Delta} \right) +1}.\label{eq:gu},
\end{equation}
with $\Phi_1$, $\Phi_2$, $p_m$, and $\Delta$ being the minimum and maximum filling factor at very low and very high pressures, the turnover pressure and the logarithmic width of the transition from low to high packing density, respectively. The transition width was experimentally found to be $\Delta =0.33$ for unidirecional compression  \citep{Guettleretal2009}. Here, we assume the compression force in collisions to be unidirectional, although the inertia of the monomers, which are forced in the direction perpendicular to the impact velocity, will generate a force. We assume this effect to be negligible, because the sideways acceleration will correspond to Poisson ratios, which are 0.15 at 1 m$s^{-1}$ and 0.3 at 100 m $s^{-1}$, respectively. Additionally the total mass producing the inertial pressure in the direction of the impact velocity corresponds to all the monomers along the diameter of the agglomerate, whereas perpendicular to the impact velocity only the monomers in radial direction contribute. 

\begin{figure}
  \center
      \includegraphics[width=.3\textwidth]{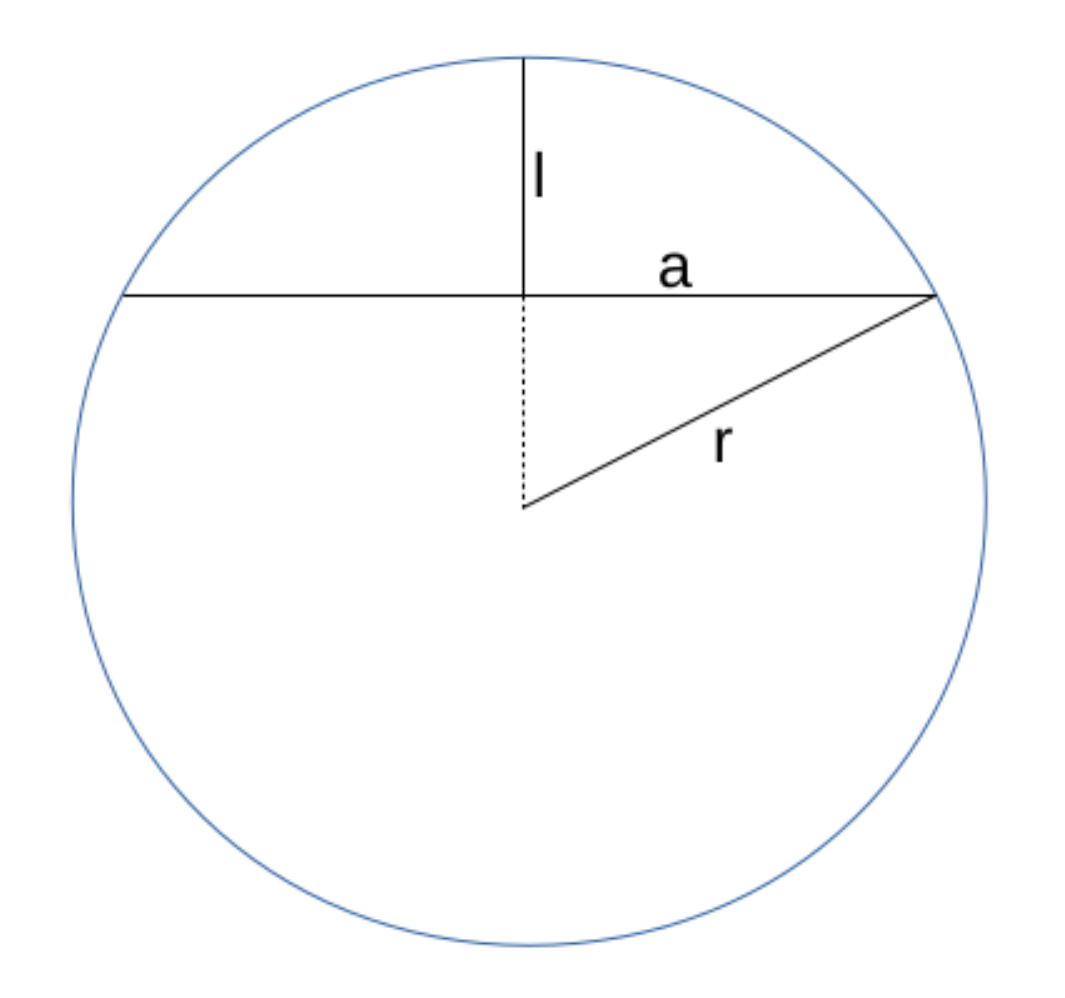}
 \caption{\label{fig:KK}Side view of a spherical cap with height $l$ and base radius $a$. The radius of the corresponding sphere is $r$. }
\end{figure}

We measured the initial packing density of our ice agglomerates to be $\Phi_1 = 0.1$. At very high pressures, the agglomerates will reach a maximum packing density for unidirectional compression of $\Phi_2 = 0.33$ \citep{BS2004}. From Eq. \ref{eq:gu}, we calculated the minimal pressure below which the agglomerates cannot be compressed, dependent on the packing density of the agglomerate, to be
\begin{equation}
 \Sigma = p_m \left(\frac{\Phi-\Phi_1}{\Phi_2-\Phi}\right)^{\Delta(\ln(10))}. \label{eq:Sigma}
 \end{equation}
Substituting the packing density with a term including the compression length $l$ leads to
 \begin{equation}
 \Phi = \frac{m}{\rho_\mathrm{S} A (h_0-l)} \label{eq:Phi},
\end{equation}
where $m$ is the mass of the colliding particles, $h_0$ their height and
\begin{equation}
 A=\left(\frac34\frac{m}{\rho_\mathrm{S} \Phi_0}\right)^\frac23 \pi^{\frac13}
\end{equation}
their cross section, where $\Phi_0=0.1$ is the initial packing density of our agglomerates. The compression force $F$ at the contact area $A_c$ of the collision partners is then calculated by assuming that the agglomerates are spheres and their contact area $A_c$ is the base of a spherical cap, i.e. 
\begin{equation}
A_c= (2 r l -l^2) \pi, 
\end{equation}
where
\begin{equation}
r=\left(\frac34\frac{m}{\pi\rho_\mathrm{S} \Phi_0}\right)^\frac13
\end{equation}
is the radius of the agglomerates (see Figure \ref{fig:KK}).

\begin{figure}
  \center
      \includegraphics[width=.8\textwidth]{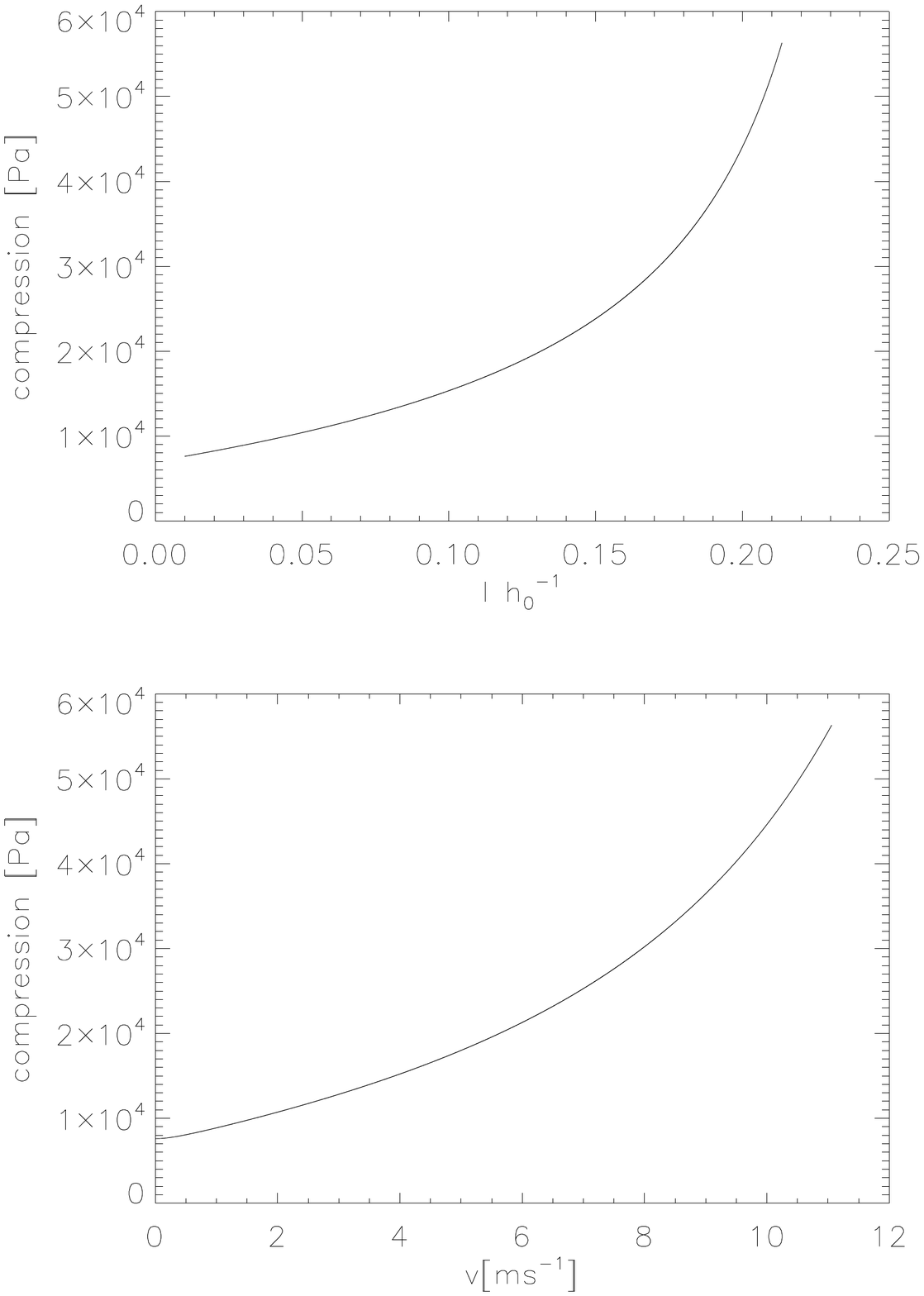}
 \caption{\label{fig:Sighv} Top: Compression stress of an agglomerate as a function of its compression length in units of the agglomerate radius. Bottom: Compressive stress of an agglomerate as a function of impact velocity.}
\end{figure}

The compression energy as a function of the deformation length is thus
\begin{equation}
 E=\int_0^{l_{max}}A_c(l) \Sigma(l) \dif l. 
 \label{eq:E}
\end{equation}
This integral was solved numerically. The impact energy is distributed to both collision partners. Therefore, $E$ occurs twice when the compression energy is compared with the kinetic energy of the impact of equal-sized partners. The relative velocity of the collision partners before the impact is therefore
\begin{equation}
  v_r=\sqrt{\frac{4E}{m_r}}, 
  \label{eq_vr}
\end{equation}
where $m_r$ is the reduced mass of the collision partners. The top graph of Figure \ref{fig:Sighv} shows $\Sigma$ as a function of the compression length $l$, normalised to the diameter of the agglomerates $h_0$. The bottom graph shows $\Sigma$ dependent on the impact velocity. $\Sigma(v)$ was calculated numerically from $v(\Sigma)$, which can be found from Equations \ref{eq:E} and \ref{eq_vr}.

The Young's modulus is the compressive stress divided by the compression length in units of the diameter of the colliding particles (Equation \ref{eq:Y}) and is defined such that the pressure is linearly dependent on the compression length. Because our compressive stress is not linearly dependent on the compression length (see Figure \ref{fig:Sighv}), this approximation results in the calculation of an averaged Young's modulus caused by the linearised pressure -- compression-length curve, i.e.
\begin{equation}
 Y=\Sigma \frac rl.\label{eq:Y}
\end{equation}
 
For each impact velocity and corresponding maximum compression, we therefore calculated a different Young's modulus. The approximation is a line in the upper graph of Figure \ref{fig:Sighv} from the origin to the given point $l h_0^{-1}$. Because higher compression leads to denser particles, the Young's modulus increases at higher impact velocities. The error done by this linearisation leads to an underestimation of the mean Young's modulus by a factor of 0.7 at low impact velocities ($<$ 1 m s$^{-1}$). For larger impact velocities, the stress--compression-length curve gets steeper and the error will increase. However, we expect fragmentation for velocities $\gg$ 1 ms$^{-1}$ where the yield concept does not apply.

\begin{figure}
  \center
      \includegraphics[width=0.5\textwidth]{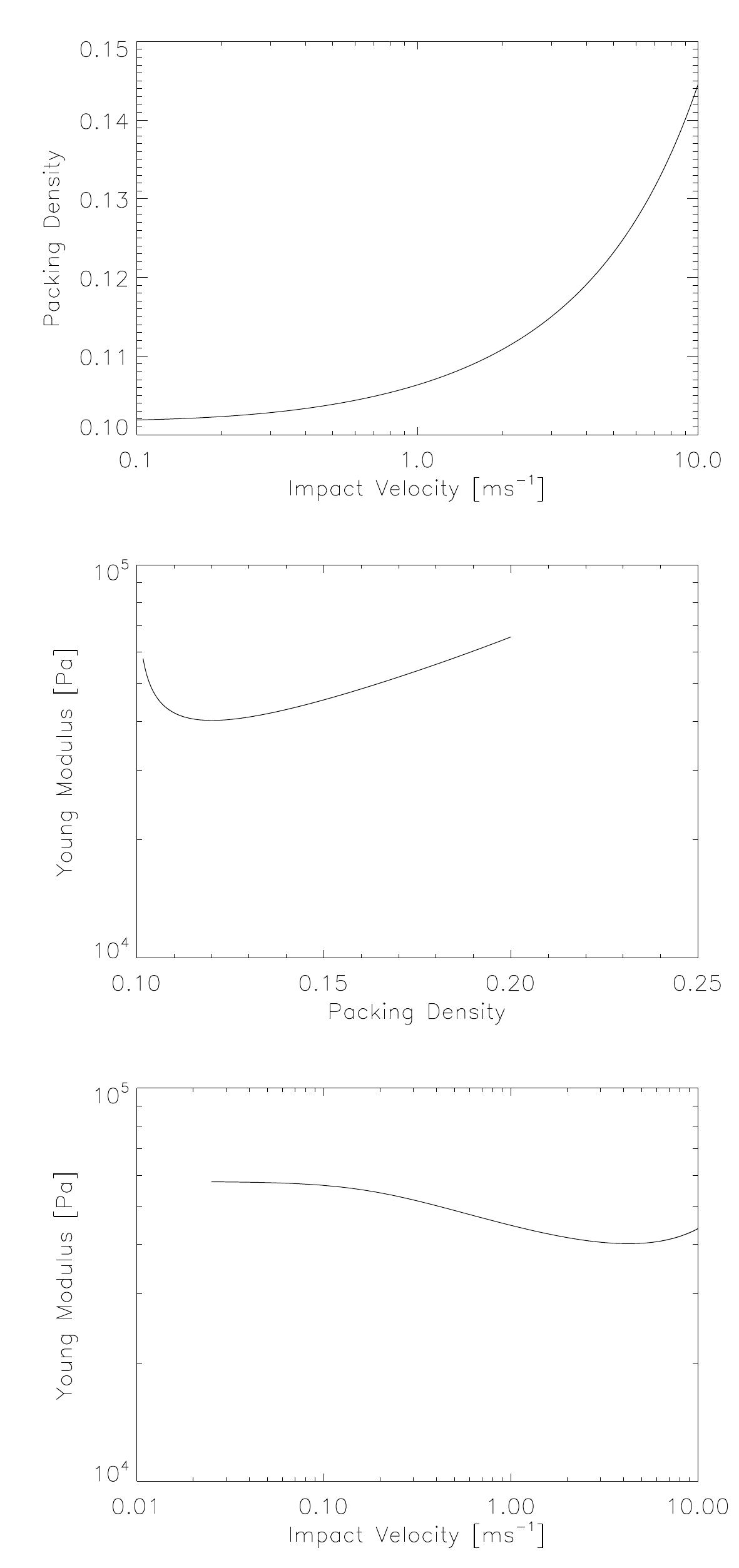}
 \caption{\label{fig:PYV}Modelling impacts between ice agglomerates. Top: The resulting packing density after a collision of ice agglomerates with an initial packing density of 0.1 and a reduced mass of $3.26\times 10^{-5}$ kg as a function of their impact velocity. centre: The Young's modulus as a function of the packing density of an ice agglomerate. Bottom: The Young's modulus as a function of the impact velocity of an ice agglomerate with an initial packing density of 0.1 and a reduced mass of $3.26\times 10^{-5}$ kg.
   }
\end{figure}

In Figure \ref{fig:PYV}, the top graph shows the packing density as a function of the impact velocity (numerically found from Eqs. \ref{eq:Phi}, \ref{eq:E} and \ref{eq_vr}). In the centre graph, the Young's modulus as a function of the packing density is plotted (calculated from Eqs. \ref{eq:Y}, \ref{eq:Phi} and \ref{eq:Sigma}). The Young's modulus' singularity at a packing density of $\Phi=0.1$ is because the packing density is 0.1 at zero pressure, which means at $l=0$. The bottom graph in Figure (\ref{fig:PYV}) shows the Young's modulus as a function of the collision velocity of the agglomerates. The curves are numerically calculated from Eqs. \ref{eq:E}, \ref{eq_vr} and \ref{eq:Y}.

\subsection{Poisson Ratio}
The Poisson ratio was derived from \citet{BS2004}. The inset of their Figure 3 shows the increase of the cross section of their high-porosity RBD sample with increasing unidirectional compression. The corresponding compression length was calculated from the packing density-vs-pressure curve (their Figure 3 and Equation \ref{eq:gu}), where the velocity-dependent pressure was calculated from the impact energy and the collisional velocity via Eqs. \ref{eq:E} and \ref{eq_vr}. This was only possible for pressures below 10$^5$Pa, because above that value, the measured cross section curve is not smooth, which cannot be treated with our method.

We fitted the cross section as a function of compressive stress from \citet{BS2004} using the  function
\begin{equation}
\frac{A}{A_1}(\Sigma)=0.125 \lg \Sigma +0.75.
\end{equation}
The radius change is then
\begin{equation}
    \frac{\Delta r}{r}=\sqrt{\frac{A}{A_1}}-1\label{eq:crc}
\end{equation}
and the compressive length change becomes
\begin{equation}
    \frac{\Delta s}{s}=\frac{\Phi_1}{\Phi} \frac{A_1}{A}.\label{eq:cls}
\end{equation}
The corresponding Poisson ratio is defined as
\begin{equation}
    \nu = \frac{\Delta r}{r} / \frac{\Delta s}{s}.
\end{equation}
The velocity is contained in $\Phi$, which is again obtained by equating the kinetic energy of the impactors with the compression energy, which can again only be found numerically because of the implicit expressions.

\begin{figure}
 \center
      \includegraphics[width=0.5\textwidth]{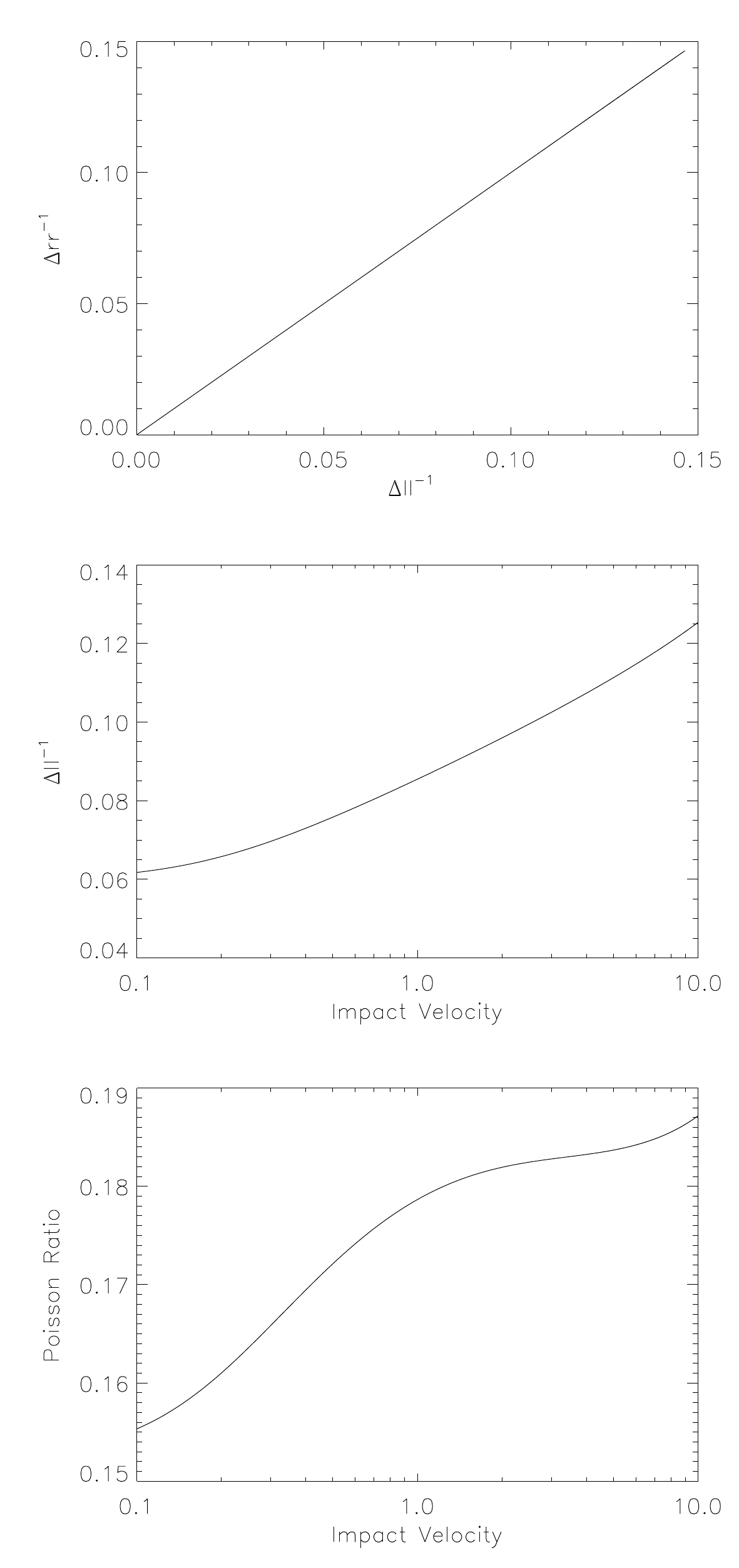}
 \caption{\label{fig:PEYV} Top: The change in compressive radius (see Equation \ref{eq:crc}) as a function of the compressive length change (see Equation \ref{eq:cls}). Centre: The change in compressive length (see Equation \ref{eq:cls}) as a function of the impact velocity. Bottom: The Poisson ratio as a function of the impact velocity.}
\end{figure}

\begin{figure}
  \center
      \includegraphics[width=0.6\textwidth]{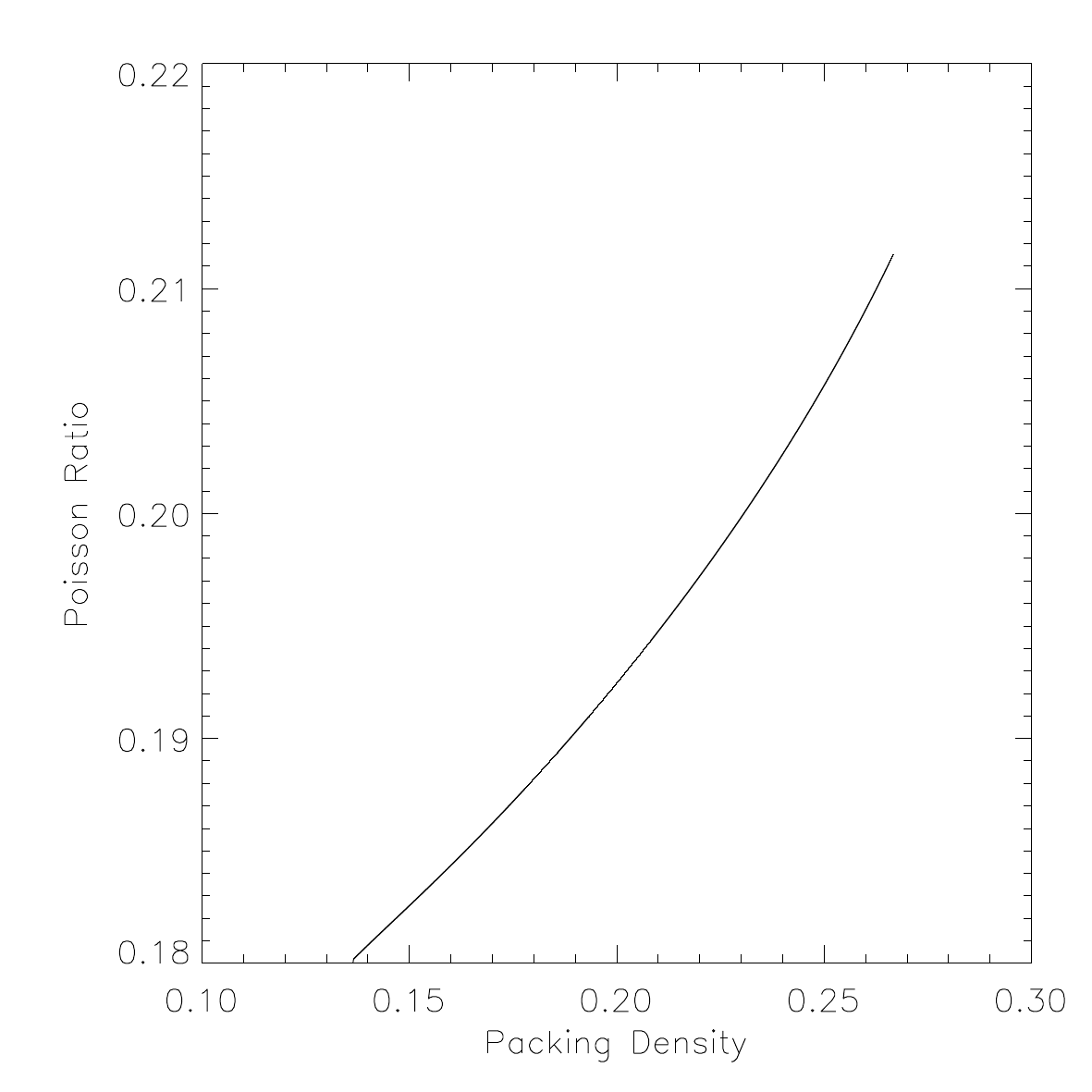}
 \caption{\label{fig:PYVX} The Poisson ratio of a high-porosity RBD ice agglomerate as a function of the packing density.  }
\end{figure}

Figs. \ref{fig:PEYV} (bottom graph) and \ref{fig:PYVX} show the Poisson ratio as a function of impact velocity and packing density, respectively.

The Poisson ratio is defined to be a constant quantity. Figure \ref{fig:PEYV} shows that the Poisson ratio does not vary considerably so that the error caused by our ansatz is acceptable.

\subsection{Viscosity}\label{sect:Vis}
There are two types of viscosity defined. The shear viscosity $\eta_1$, which occurs in a shear flow, and the volume viscosity $\eta_2$, which occurs at a compressive or expansive flow. The shear viscosity is defined  by the stress $\sigma$ divided by the shear rate $\dot{\gamma}$, i.e.
\begin{equation}
\eta_1=\frac{\sigma}{\dot{\gamma}}.
\end{equation}
In our model, the stress is given by $\Sigma(v)$. We calculated the shear rate from
\begin{equation}
    \dot{\gamma}=\frac{v_{r}}{\sqrt{2}r}. \label{eq:shear}\\
\end{equation}
The shear radius $r_s$ is zero at the contact point and $r$ at the equator of the particle where the mean shear radius is $r~ 2^{-0.5}$, which us used in Equation \ref{eq:shear}. The factor $2$ occurs, because the shear rate for each of the collision partners corresponds to half of their relative velocity. The contact radius of the agglomerates at maximum compression is $r_c$. In other words, the particle is stopped at the contact point and the material around that point is moving further on and causes shear. We assumed a linear velocity gradient (ansatz for shear viscosity).

The volume viscosity is defined as the stress $\sigma$ times the volume change over time relative to the initial volume, i.e.
\begin{equation}
\eta_2=\sigma\frac{\dot{V}}{V}.
\end{equation}

The volume change by an impact in z-direction would be constant over the cylinder height in the case a cylinder would impact along its axis. In our case, spheres collide and we therefore have to use the average cross section in impact direction. The ratio that has to be taken into account is
\begin{equation}
    d=\frac{r^2 \pi r}{\frac43 r^3 \pi}=\frac34.\label{eq:corr}
\end{equation}

With the above considerations, the volume change can be expressed by the impact velocity, using the continuity equation
\begin{equation}
    \dot{\rho} +\rho \mbox{ div } \vec{v}=0\nonumber 
\end{equation}
\begin{equation}
    \mbox{div } \vec {v}=\frac{\dot{\rho}}{\rho}.
\end{equation}
Because the agglomerate mass is constant, this can be written as
\begin{equation}
    \mbox{div } \vec{v}=-\frac{V}{\dot{V}}.
\end{equation}

Replacing $\mbox{div }\vec{v}$ by the z-component $v$ and applying our correction Equation \ref{eq:corr}, we finally get
\begin{equation}
    \eta_2=\frac43 \frac{\sigma}{v}.
\end{equation}
Because in our simple calculation
\begin{equation}
    \frac34 \approx \frac{1}{\sqrt{2}}
\end{equation}
is true,
\begin{equation}
    \eta_1 \approx \eta_2
\end{equation}
is also true. Which holds for many compressible materials for which the volume viscosity is based on the same physical processes as the shear viscosity. This is the case for our granular material, because both viscosities are based on rolling and sliding forces of the particles the material consists of. The stress value is again taken from Equation \ref{eq:Sigma} for a $\Phi$ at a compression state given from the impact velocity. We used this final value as an approximation, because it was not possible to follow the compression during the impact in the model of \citet{Brilliantov2005}.

\begin{figure}
  \center
      \includegraphics[width=.5\textwidth]{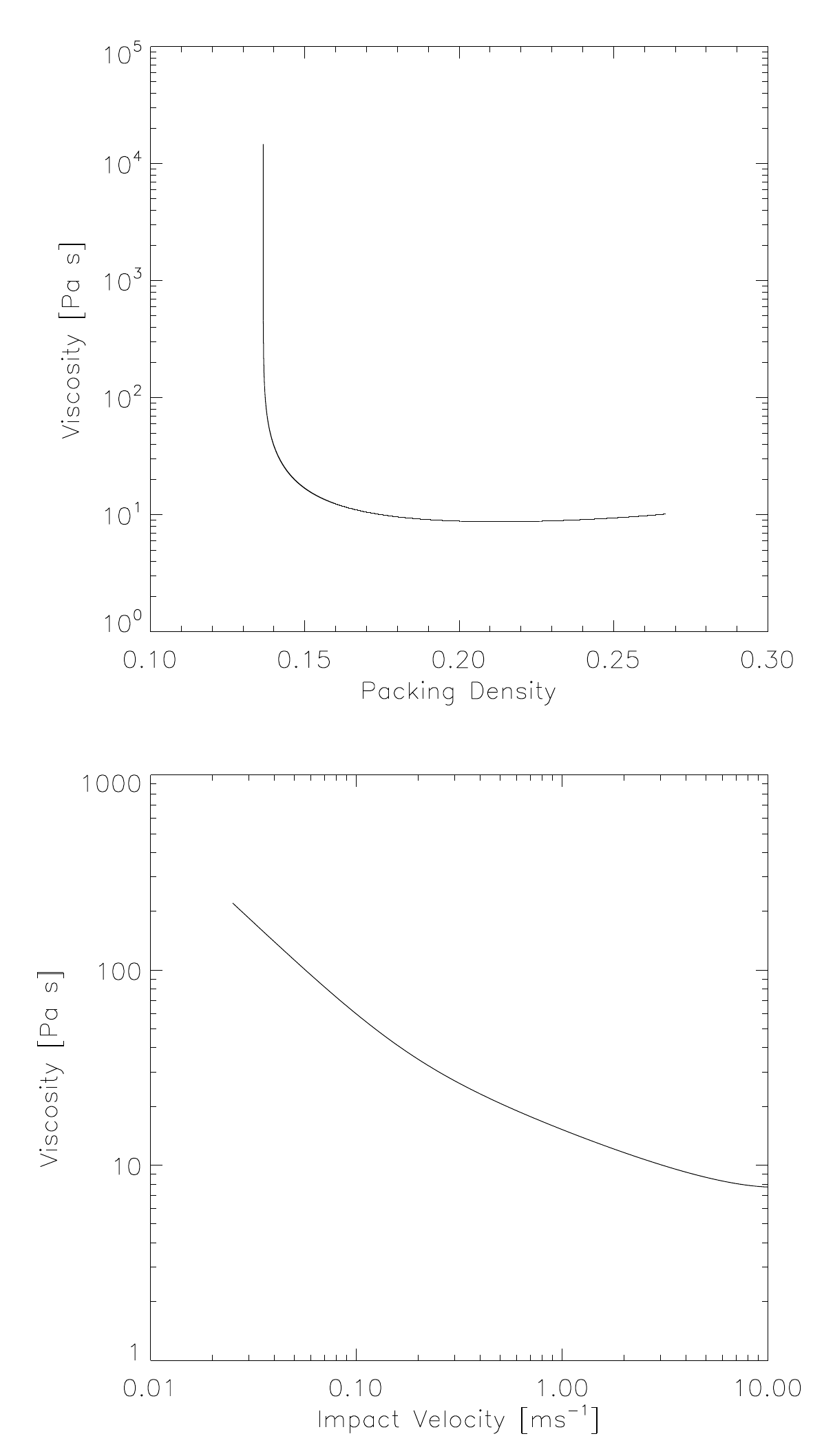}
 \caption{\label{fig:eta} Top: The viscosity of an agglomerate as a function of its packing density. Bottom: The viscosity as a function of impact velocity.
   }
\end{figure}

Figure \ref{fig:eta} shows the viscosity of an agglomerate as a function of packing density (top graph) and impact velocity (bottom graph) in an impact event. The packing density corresponds to the impact compression of the agglomerate. With decreasing impact velocity, the corresponding compression gets so low that it cannot overcome the rolling friction in the agglomerate. For these velocities, the agglomerate behaves like a solid-state body and its viscosity increases dramatically. In other words, the slope of the volume filling factor as a function of pressure gets very flat at low stresses (see \citet{BS2004}, their Figure 3). At packing densities larger than 0.2, the viscosity increases again slightly, because the coordination number of the agglomerate increases, which causes higher friction during deformation.

\subsection{Coefficient of Restitution}\label{sect:CR}
In the following, we use the velocity-dependent Young's modulus, viscosity, packing density and Poisson ratio, calculated in the precious sections, to numerically derive the coefficient of restitution, again applying the model of \citet{Brilliantov2005} that manifests in their Equation 27. We used the parameters given in Table \ref{table:Para2} for our calculations. The coefficient of restitution we get for the ice agglomerates, applying the surface energy from \citet{Gundlachetal2018} (20 mJ m$^{-2}$), is plotted in Figure \ref{fig:RCIDE} with a dotted curve, whereas the dashed curve shows the calculated coefficient of restitution with the surface energy of \citet{Makkonen1997} (77 mJ m$^{-2}$) applied.

However, note that \citet{Brilliantov2005} assumed compact particles. Therefore, we reduced the surface energy as a function of the packing density according to
\begin{equation}
    \gamma_{eff} = \gamma \Phi^\frac23. \label{eq:gampd}
\end{equation}
We used the $p_m$ of omnidirectional compression from \citet{Rezaei} for our ice particles, because unidirectional compression data for ice agglomerates do not exist.
Therefore, the real value is expected to be smaller. But it is a better choice than to use the unidirctional SiO$_2$ data from \citep{Guettleretal2009}.
We used the surface energy of ice found by \citet{Makkonen1997} for the coefficient of restitution, because it fits our ice agglomerates best. 

Our model cannot reproduce the Bingham behaviour found in our experiments. We think this is because we used the impact model of \citet{Brilliantov2005}, which itself uses constants for the Young's modulus, Poisson ratio and viscosity. However, these values certainly depend on the packing density of the agglomerates, which changes during impact. Instead, we calculated and implemented these parameters for the final compression of our agglomerates, which depends on the impact velocity and the agglomerate masses. This is a good approximation for the Young's modulus and the Poisson ratio, but a poor approximation for the viscosity, which is a main source of energy dissipation and therefore connected to the Bingham behaviour.

In Figure (\ref{fig:RCDDE}), the surface energy of \citet{Heim1999} was used (see Table \ref{table:Para2}) to model the behaviour of our dust agglomerate in the impact experiments, where $\Delta$ was taken from \citet{Guettleretal2009}. We again had to reduce it by the packing density according to Equation \ref{eq:gampd}. The sticking threshold velocity is about a factor of 5 lower than in our experiments. We think this is due to the fact that for the calculation we used spherical particles whereas the experimental particles were cuboids.

\section{\label{APP:BH}Illustration of the Bingham behaviour of our particles}
 
This Section is not intended to calculate the exact velocity at which the Bingham behaviour occurs in our experiment, but is meant to derive a simple order-of-magnitude estimation to illustrate the mechanism of the Bingham behaviour. To achieve this, we use the physical properties of the water-ice agglomerates. The Bingham behaviour occurs in a region in which the rolling and sliding thresholds are not exceeded so that the agglomerate can only be deformed elastically.

\subsection{Rolling threshold}

In the work of \citep{Gundlachetal2011}, the rolling threshold $F_r$ was measured for ice particles with 1.45 $\mu$m radius to be $(1.15 \pm 0.24) \times 10^{-8}$ N. 
Because the sliding threshold force is larger  than the rolling threshold  \citep{DominikTielens1997} it is not considered here.

\subsection{\label{sect:def} Deformation threshold force on ice agglomerates}

For this estimation, we assume cylinder-shaped agglomerates with a cylinder radius of $r_c = 0.005$ m and a height of $h_c = 0.01$ m. Thus, the total force such a agglomerate can withstand is
\begin{equation}
    F_t = F_r \frac{r_c^2 \pi}{r_m^2 \pi \Phi} \approx 0.34 \mbox{N}.
\end{equation}

\subsection{\label{sect:dist}Distance of elastic rolling and maximum deceleration}

To derive the distance of elastic rolling and maximum deceleration, we follow the assumption of \citet{DominikTielens1997} to be one atom diameter ($l_{er}=10^{-10}$ m).

Thus, the length of deceleration of the ice agglomerates is
\begin{equation}
    l_t=l_{er} \frac{h_c}{r_m} \approx 3.4 \times 10^{-7} m.
\end{equation}
The maximum deceleration that the agglomerates can withstand without plastic deformation is
\begin{equation}
    a_t=\frac{F_t}{m}\approx 4360 \frac{m}{s^2},
\end{equation}
with $m_c=r_c^2 \pi h_c \rho_{ice} \Phi_{ice} \approx 7.8 \times 10^{-5}$ kg. Here, $\rho_{ice}$ is the mass density of water ice (see Table \ref{table:Para}) and $\Phi_{ice}$ is the initial packing density of the ice agglomerates (see Table \ref{table:Para2}).

\subsection{Maximum velocity at which the Bingham behaviour can occur}
From sections \ref{sect:def} and \ref{sect:dist}, the maximum velocity at which the Bingham behaviour can occur can be calculated to
\begin{equation}
     v_B=\sqrt{2 a_t l_t} \approx 0.04 \frac{m}{s}.
\end{equation}
In case we use the experimental value of $l_{er}=32 \times 10^{-10}$ m from \citet{Heim1999}, the maximum velocity for the Bingham behaviour is
\begin{equation}
      v_B \approx 0.22 \frac{m}{s}.\label{eq:hwe}
\end{equation}


\end{document}